\documentclass[pdflatex,sn-nature]{sn-jnl}

\usepackage{geometry}
\usepackage{etoolbox}
\makeatletter
\patchcmd{\@outputpage@head}{\@tempdima}{0pt}{}{}
\makeatother
\makeatletter
\@twosidefalse  
\makeatother

\usepackage{graphicx}%
\usepackage{multirow}%
\usepackage{amsmath,amssymb,amsfonts}%
\usepackage{amsthm}%
\usepackage{mathrsfs}%
\usepackage[title]{appendix}%
\usepackage{xcolor}%
\usepackage{textcomp}%
\usepackage{manyfoot}%
\usepackage{booktabs}%
\usepackage{algorithm}%
\usepackage{algorithmicx}%
\usepackage{algpseudocode}%
\usepackage{listings}%
\usepackage{quantikz}
\usepackage{braket}
\usepackage[caption=false]{subfig}
\usepackage{stmaryrd}
\usepackage[T1]{fontenc}

\begin{document}

\title[Article Title]{\textbf{Scaling of silicon spin qubits under correlated noise}}

\author*[1]{{Juan} {S.} {Rojas-Arias}}\email{juan.rojasarias@riken.jp}

\author*[2]{{Leon} {C.} {Camenzind}}\email{leon.camenzind@riken.jp}

\author[2]{{Yi-Hsien} {Wu}}

\author[1,3]{{Peter} {Stano}}

\author[2]{{Akito} {Noiri}}

\author[2]{{Kenta} {Takeda}}

\author[2]{{Takashi} {Nakajima}}

\author[1]{{Takashi} {Kobayashi}}

\author[4]{{Giordano} {Scappucci}}

\author[1,5,6,7,8]{{Daniel} {Loss}}

\author*[1,2]{{Seigo} {Tarucha}}\email{tarucha@riken.jp}

\affil[1]{\orgdiv{RIKEN Center for Quantum Computing (RQC)}, \orgname{RIKEN}, \orgaddress{\street{Hirosawa 2-1}, \city{Wako-shi}, \postcode{351-0198}, \state{Saitama}, \country{Japan}}}

\affil[2]{\orgdiv{RIKEN Center for Emergent Matter Science (CEMS)}, \orgname{RIKEN}, \orgaddress{\street{Hirosawa 2-1}, \city{Wako-shi}, \postcode{351-0198}, \state{Saitama}, \country{Japan}}}

\affil[3]{\orgdiv{Slovak Academy of Sciences}, \orgname{Institute of Physics}, \orgaddress{\street{845 11}, \city{Bratislava}, \country{Slovakia}}}

\affil[4]{\orgdiv{QuTech and Kavli Institute of Nanoscience}, \orgname{Delft University of Technology} \orgaddress{\street{Lorentzweg 1}, \city{Delft}, \postcode{2628 CJ}, \country{Netherlands}}}

\affil[5]{\orgdiv{Department of Physics}, \orgname{University of Basel}, \orgaddress{\street{Klingelbergstrasse 82}, \postcode{CH-4056}, \city{Basel}, \country{Switzerland}}}

\affil[6]{\orgdiv{Physics Department}, \orgname{King Fahd University of Petroleum and Minerals}, \postcode{31261}, \city{Dhahran}, \country{Saudi Arabia}}

\affil[7]{\orgdiv{Quantum Center}, \orgname{KFUPM}, \city{Dhahran}, \country{Saudi Arabia}}

\affil[8]{\orgdiv{RDIA Chair in Quantum Computing}}

\abstract{The path to fault-tolerant quantum computing hinges on hardware that scales while remaining compatible with quantum error correction (QEC). Silicon spin qubits are a leading hardware candidate because they combine industrial fabrication compatibility with a nanoscale footprint that could accommodate millions of qubits on a chip. However, their suitability for QEC remains uncertain since spatially correlated noise naturally emerges from the resulting close proximity of qubits. These correlations increase the likelihood of simultaneous errors and erode the redundancy that QEC depends on.
Here we quantify the spatial extent of noise correlations in a five-qubit silicon array and assess their impact on QEC. We identify two distinct sources of correlated noise: global magnetic field drifts that generate perfectly correlated fluctuations, and charge noise from two-level fluctuators that produces short-range correlations decaying within neighboring qubits. While magnetic drifts represent a critical correlated noise source that can compromise QEC, they can be mitigated. In contrast, the measured charge noise correlations are moderate, electrically tunable, and compatible with fault-tolerant operation with minimal qubit overhead.
Our results establish quantitative benchmarks for correlated noise and clarify how such correlations impact the viability of quantum error correction in scalable qubit arrays.
}

\maketitle

The realization of a large-scale quantum computer requires protection of quantum information from errors induced by noise and imperfections. Quantum error correction (QEC) provides the framework for achieving this goal, enabling fault-tolerant computation beyond the limits imposed by physical error rates \cite{NielsenChuang}. By encoding logical qubits into entangled states of many physical qubits, QEC enables detection and correction of errors without collapsing the encoded information. The \textit{threshold theorem} guarantees that, provided physical errors remain below a certain threshold, arbitrarily low logical error rates can be achieved by increasing the code distance, that is, by increasing the number of physical qubits used to encode a logical qubit \cite{Aharonov1997,Knill1998,Kitaev1997}. Recent experiments with superconducting qubits \cite{Google2023}, bosonic qubits \cite{Putterman2025}, trapped ions \cite{Hong2024}, and neutral atoms \cite{Bluvstein2025} have demonstrated this principle in quantum hardware, even achieving logical error rates below those of the underlying physical qubits \cite{Google2025}. 

In its most basic form, the threshold theorem assumes that errors are temporally and spatially uncorrelated \cite{Preskill1998}. While QEC can tolerate certain forms of temporal correlations \cite{Hutter2014,Bombin2016}, spatial correlations can be far more problematic \cite{Klesse2005,Clader2021}, particularly when they are long-range \cite{Clemens2004,Aharonov2006,Terhal2005}. Correlated errors acting simultaneously on multiple qubits reduce the effective redundancy of the code and thereby its efficiency in suppressing logical errors as the code distance increases \cite{Klesse2005}; in some regimes, they can even invalidate the threshold entirely \cite{Clemens2004}. Indeed, correlated errors have already emerged as a limiting factor in recent superconducting-qubit demonstrations of QEC \cite{Google2023,Google2025}.

Spin qubits are a leading platform for large-scale quantum processors, combining high-fidelity control with a compact footprint that is compatible with standard semiconductor fabrication. This compatibility with mature semiconductor technology provides a realistic route toward the million-qubit scales required for implementing fault-tolerant quantum computing with QEC codes \cite{Gidney2021,Katabarwa2024,Gidney2025}. Moreover, these qubits already support high-fidelity operation, with single-qubit gate infidelities approaching $10^{-5}$ \cite{Wu2025,Stano2022} and two-qubit gate infidelities approaching $10^{-3}$ \cite{Noiri2022,Mills2022,Xue2022,Madzik2022}. However, the same compact footprint that makes spin qubits attractive for scalability also increases their susceptibility to spatially correlated noise \cite{Yoneda2023,Rojas-Arias2023}. Thus, the compatibility of spin qubits with quantum error correction becomes a critical open question.

We explore this question here, investigating noise correlations in a five-qubit silicon array. We quantify them, analyze their spatial scaling using a microscopic two-level fluctuator model, and demonstrate their electric tunability. In addition, using analytics and numerical simulations, we evaluate their impact on logical error rates and resource overhead in conventional error correction protocols. Our results reveal two dominant sources of correlated noise: global magnetic field drifts and charge noise. The former can critically limit QEC performance, but it is technical in origin and has clear mitigation strategies. The latter produces short-range correlations that remain compatible with fault-tolerant operation. Our main conclusion is that correlated noise in silicon spin qubits, while present, does not represent a fundamental obstacle to fault-tolerant quantum computing at the levels observed here. Instead, our work establishes quantitative benchmarks for correlated noise and provides a general framework for assessing its impact on quantum error correction.

\section*{Correlated noise regimes}

\begin{figure}
\centering
\subfloat{\includegraphics[width=\textwidth]{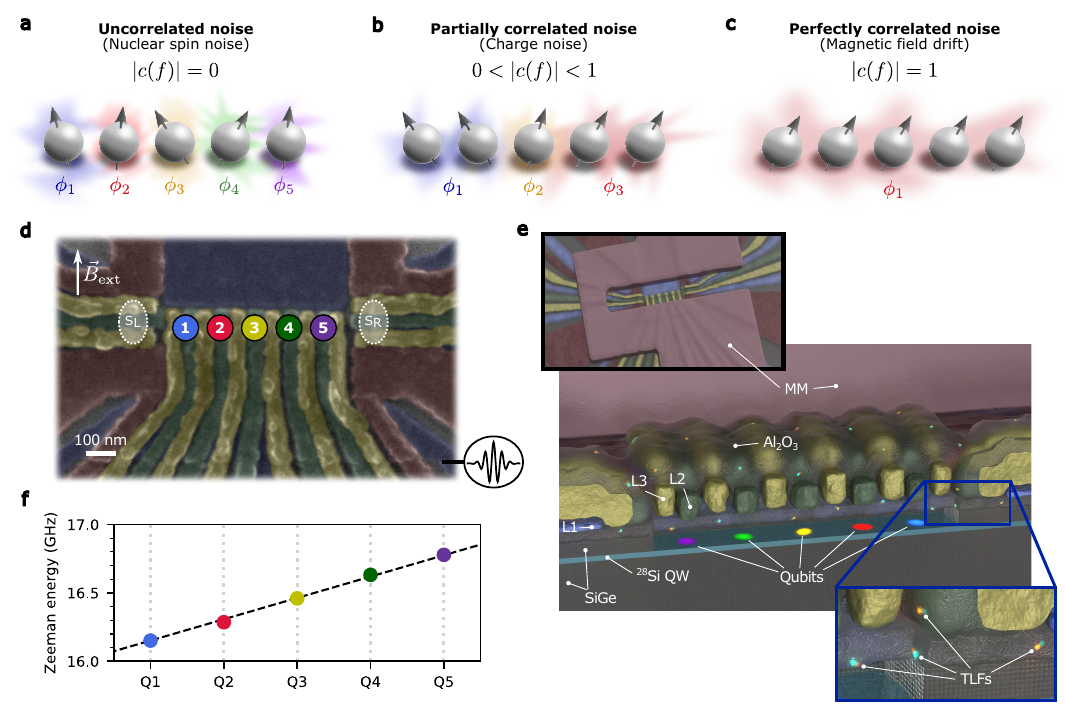}\label{fig:device_a}}
\subfloat{\label{fig:device_b}}
\subfloat{\label{fig:device_c}}
\subfloat{\label{fig:device_d}}
\subfloat{\label{fig:device_e}}
\subfloat{\label{fig:device_f}}
\caption{\textbf{Noise correlation regimes and device layout}. 
(a-c) Schematic illustrations of qubits subject to dephasing noise in different correlation regimes: (a) uncorrelated, (b) partially correlated, and (c) perfectly correlated. Qubits are represented as spheres with arrows accumulating random phases $\phi_i$, where the degree of correlation determines whether these phases are independent, partially shared, or identical. 
(d) Scanning electron microscope image of a device nominally identical to the one used in the experiment. Colored circles indicate the location of the five qubits, labeled Q1--Q5. Charge sensors $S_L$ and $S_R$ (white ovals) are used for spin-to-charge conversion and readout. An external in-plane magnetic field $\vec{B}_\mathrm{ext}$ is applied perpendicular to the qubit array. Global microwave control enables single-qubit rotations. 
(e) 3D rendering of the device structure. The $^{28}$Si quantum well hosting the qubits is shown in light blue. Charge noise arises from two-level fluctuators in the surrounding oxide layers and couples to the spins via the artificial spin-orbit field generated by a cobalt micromagnet (purple). (f) Measured individual Zeeman splittings for each qubit, showing the engineered energy gradient across the array. This linear gradient provides qubit addressability, allows the relative qubit positions to be inferred from their Zeeman energies, and sets the sensitivity to charge-noise-induced dephasing.}
\label{fig:device}
\end{figure}

To assess the role of correlated noise in QEC, we first classify the relevant correlation regimes and relate them to distinct physical sources in our spin-qubit platform. Since here the energy relaxation time can be made very long \cite{Koch2025}, we focus on dephasing noise as the dominant noise channel, which appears as temporal fluctuations in the spins' Zeeman energies.

We quantify correlations between a qubit pair using the normalized cross-power spectral density (cross-PSD) $c(f)$ (see Methods). This dimensionless measure ranges from 0 to 1 in magnitude and enables us to distinguish three characteristic regimes (Fig.~\ref{fig:device_a}-\ref{fig:device_c}):

\textbf{\textit{Uncorrelated noise}} ($|c(f)| = 0$; Fig.~\ref{fig:device_a}): This regime corresponds to independent fluctuations in each qubit and represents the most favorable scenario for QEC, serving as the standard reference case in threshold analyses. In spin qubits, a typical source of uncorrelated noise is hyperfine coupling to nearby nuclear spins \cite{Rojas-Arias2026a,Reilly2008,Malinowski2017}. Dominated by the contact hyperfine interaction, the resulting noise is local \cite{Yoneda2023,Rojas-Arias2023}. The diffusion of nuclear spins \cite{Hayashi2008} is too slow to impose an appreciable long-range coupling over experimentally relevant timescales, typically on the order of hours.

\textbf{\textit{Perfectly correlated noise}} ($|c(f)| = 1$; Fig.~\ref{fig:device_c}): In this regime, fluctuations in the qubits' phases are identical in time, increasing the likelihood of correlated errors that severely degrade QEC performance. Such correlations arise from global noise sources that couple to all qubits, such as magnetic field drifts \cite{Muhonen2014} or voltage fluctuations on shared control gates. These correlations persist irrespective of interqubit distance, raising concerns for scalable architectures. On the other hand, the perfect symmetry of this noise enables mitigation strategies, such as encoding into decoherence-free subspaces \cite{Burkard2023}.

\textbf{\textit{Partially correlated noise}} ($0 < |c(f)| < 1$; Fig.~\ref{fig:device_b}): This intermediate regime poses the greatest challenge. Correlations may be strong enough to undermine QEC, yet too weak or spatially variable to exploit decoherence-free subspaces. An example is charge noise, which couples to spin qubits via spin-orbit interactions \cite{Camenzind2022,Piot2022} or magnetic field gradients \cite{Yoneda2023}, being the case here. In semiconductor devices, charge noise is usually attributed to ensembles of two-level fluctuators (TLFs) located in oxide layers of the heterostructure \cite{Kuhlmann2013,Connors2019,Ye2024,Rojas-Arias2026} (see Fig.~\ref{fig:device_e}), resulting in distance- and frequency-dependent correlations between qubits \cite{Yoneda2023,Donnelly2025,Rojas-Arias2023,Rojas-Arias2026a}. The diversity and microscopic nature of charge-noise make its correlated effects highly device- and configuration-dependent, complicating the development of general mitigation strategies.

Together, these regimes highlight that the spatial structure of noise correlations is a key determinant of QEC performance. Remarkably, spin qubits host noise sources for each of the above categories to an excellent approximation. Understanding the spatial extent and physical origin of correlated noise in spin-qubit arrays is therefore critical for assessing their scalability, and is the central question addressed in this work.

\section*{Noise in a silicon five-qubit device}

To investigate noise correlations along these lines, we use data from a five-qubit device fabricated on a Si/SiGe heterostructure with an isotopically enriched $^{28}$Si quantum well (Figs.~\ref{fig:device_d} and \ref{fig:device_e}; see Methods for fabrication and device details). An in-plane magnetic field is applied perpendicular to the array, and a patterned Co micromagnet provides a magnetic gradient that enables both electrical spin control and qubit-specific Zeeman splittings (Fig.~\ref{fig:device_f}). The device is operated in a regime where exchange interactions between qubits are negligible ($<10$~kHz), ensuring that measured correlations originate from the common environment rather than direct inter-qubit couplings.

\begin{figure}
\centering
\subfloat{\includegraphics[width=\textwidth]{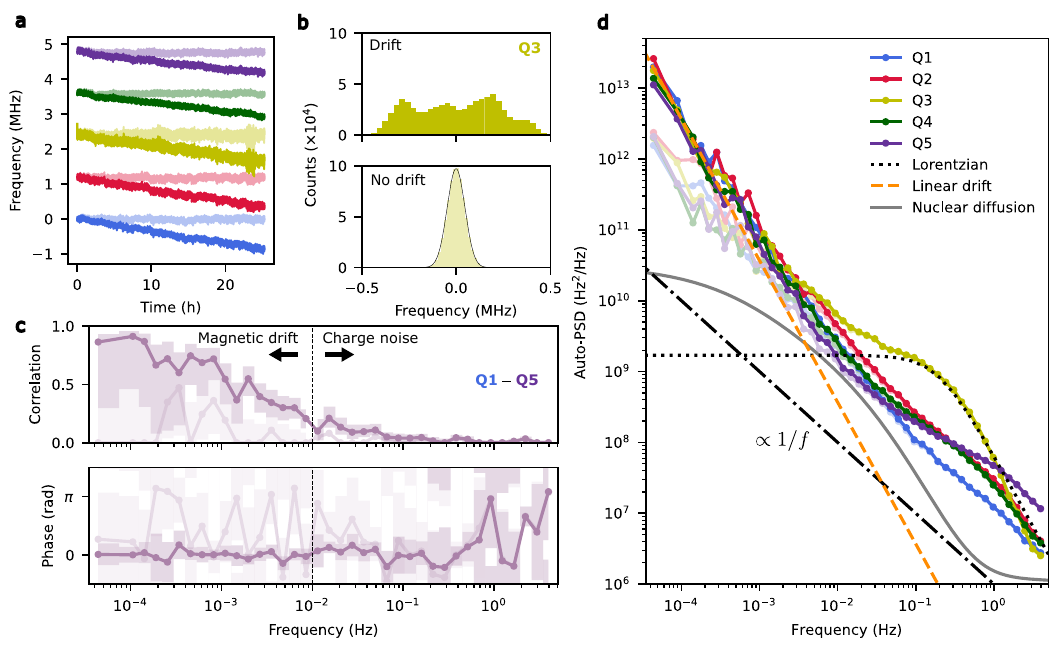}\label{fig:drift_a}}
\subfloat{\label{fig:drift_b}}
\subfloat{\label{fig:drift_corr}}
\subfloat{\label{fig:auto-PSD}}
\caption{\textbf{Separation of global magnetic drift and local charge noise.} 
(a) Measured time traces of the qubit energies over a 24-hour period, showing a clear negative drift attributed to a gradual loss of magnetization in the superconducting magnet. Lighter traces show the same data after subtracting a linear drift from each qubit individually. 
(b) Histograms of the energy fluctuations of qubit Q3 before (top) and after (bottom) drift removal. The drift causes a skewed distribution, which reverts to a Gaussian distribution after the removal. 
(c) Normalized cross-PSD between the most distant qubit pair, Q1--Q5. The upper panel shows the magnitude, and the lower panel the phase. Strong low-frequency correlations with phase around 0 are observed, consistent with a global magnetic field drift. After drift removal (faint traces), the magnitude drops and the phase is no longer locked at a single value, indicating uncorrelated noise. Occasional spikes of apparent correlation at low frequencies are attributed to the limitations of the simplistic linear drift removal and reduced statistical confidence in spectral estimates at these frequencies. A vertical dotted line marks a crossover, which takes place somewhere around $10^{-2}$~Hz, judged from the auto-PSDs in (d) transitioning from a uniform, drift-dominated regime to qubit-specific spectra with charge-noise features.
(d) Auto-PSDs for all five qubits. Drift removal (faint colors) reduces the low-frequency noise by an order of magnitude. Above $10^{-2}$~{Hz}, the spectra become qubit-specific. Q3 exhibits a prominent Lorentzian peak, consistent with coupling to a TLF with a characteristic switching time of $0.7$~s, shown as a dotted-line fit. The dashed orange line shows the expected auto-PSD for a linear drift of 8~Hz/s, calculated using Eq.~\eqref{eq:drift_psd} (Methods). The solid gray line shows the predicted noise spectrum from nuclear spin diffusion, including oscillations of the electron wavefunction arising from the valley degree of freedom in silicon \cite{Rojas-Arias2026a}. The measured spectra exceed the nuclear spin noise prediction, indicating that nuclear spin diffusion is not the dominant dephasing mechanism in our device, although a finite contribution from nuclear spins cannot be entirely excluded. The black dash-dotted line presents a reference $1/f$ power-law dependence.}
\label{fig:drift}
\end{figure}

We monitor qubit-energy fluctuations via repeated interleaved Ramsey sequences continuously performed over $\sim24$ hours (see Methods and Extended Data Fig.~\ref{exfig:sequence} for sequence details). The resulting time traces (Fig.~\ref{fig:drift_a}) reveal a global downward drift in qubit energies at an average rate of $\sim8$~Hz/s, which we attribute to gradual demagnetization of the superconducting magnet \cite{Muhonen2014}.  Because a static magnetic field is an essential ingredient of spin-qubit platforms, such low-frequency magnetic drifts represent a generic technical challenge. {Interestingly, the drift rate varies across qubits, the steepest being for Q1 and approximately 50\% weaker for Q5. The origin of this variation remains unclear and may reflect the combined effect of the global magnetic-field drift with additional ultra-low-frequency fluctuations, or changes in the magnetization of the Co micromagnet with the external field.} These energy shifts correspond to a drift of $\sim3\times10^{-7}$~mT/s, which is within the specification of the magnet. Systematic tests across multiple cryogenic setups show that this behavior appears consistently in several superconducting magnets, while being absent in others (see Methods). Whenever present, such drifts constitute a significant source of spatially uniform, distance-independent correlated noise.

Figure~\ref{fig:drift_corr} illustrates this effect via the normalized cross-PSD $c_{1,5}(f)$ between the most distant qubits, Q1 and Q5. At the lowest frequencies ($f \leq 10^{-3}$~Hz), the qubits exhibit near-perfect correlation, a behavior consistent across all qubit pairs (see {Extended Data Fig.~\ref{exfig:full_matrix}}). Correlations diminish gradually with increasing frequency, vanishing beyond approximately $10^{-1}$~Hz. Since global magnetic fluctuations yield perfect correlations at all frequencies, the observed decay indicates the onset of a distinct noise source dominating above roughly $10^{-2}$~Hz (dashed line).

To isolate this secondary noise contribution, we remove the magnet drift by fitting and subtracting a linear trend from each qubit's energy trace (light colors in Fig.~\ref{fig:drift_a}). The histograms in Fig.~\ref{fig:drift_b} show that drift correction restores Gaussian statistics to the distribution of qubit-energy fluctuations, confirming that the apparent non-Gaussian behavior originated from slow global drifts rather than intrinsic noise. The recalculated cross-PSD $c_{1,5}(f)$ (light colors in Fig.~\ref{fig:drift_corr}) reveals that noise correlations between distant qubits vanish.  Crucially, the phase of the correlation shifts from zero {(indicative of in-phase correlations, as expected for a global noise source)} before drift removal to a random distribution afterward, a hallmark of uncorrelated noise. The random correlation phases are also observed at high frequencies (both with or without the drift correction), where the correlation amplitude approaches zero.

The data in Fig.~\ref{fig:auto-PSD} suggest that the dominant contribution at frequencies above $10^{-2}$~Hz arises from charge noise. The auto-PSDs are, first, above the theoretical estimate of nuclear spin noise and, second, qubit-specific at higher frequencies \cite{Rojas-Arias2026a}. In particular, qubit Q3 exhibits a pronounced Lorentzian component, consistent with coupling to a TLF. Such components are a well-established signature of charge noise in isotopically purified silicon devices with micromagnets \cite{Rojas-Arias2023,Rojas-Arias2026}. Additional evidence is provided by Extended Data Fig.~\ref{exfig:full_matrix}, which shows that correlations in this frequency range depend on the qubit pair. By contrast, the low-frequency spectra in Fig.~\ref{fig:auto-PSD} display a uniform profile across all qubits, which matches our expectation for a global drift (see Methods). Drift removal leaves the high-frequency charge noise largely unaffected but suppresses the low-frequency noise power by an order of magnitude, underscoring the substantial impact of magnet drift in our data.

\section*{Control of noise correlations}

\begin{figure}
\centering
\subfloat{\includegraphics[width=\textwidth]{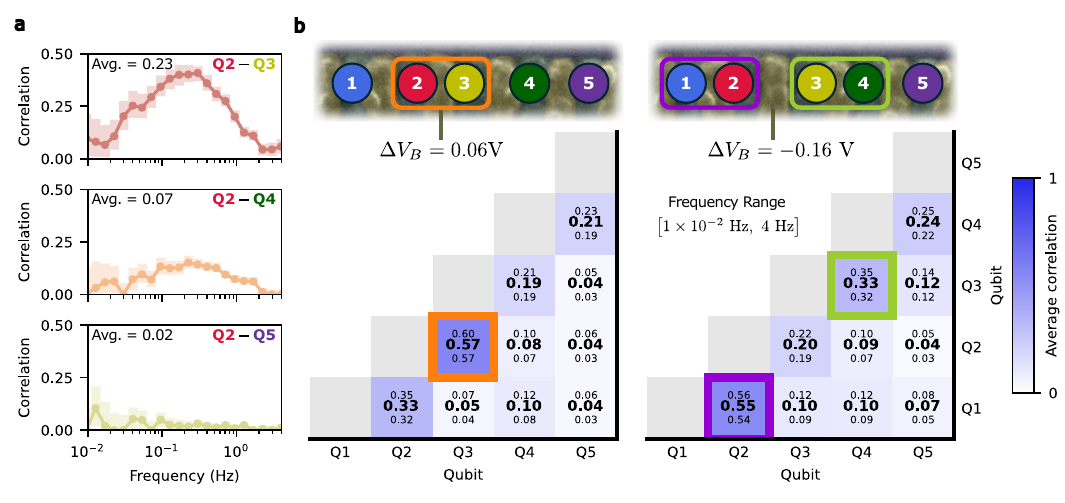}\label{fig:correlation_a}}
\subfloat{\label{fig:correlation_b}}
\caption{\textbf{Spatial correlations and their tunability by gates.} 
(a) Normalized cross-PSD magnitudes between Q2 and increasingly distant qubits: Q3 (top), Q4 (middle), and Q5 (bottom). A pronounced correlation peak is observed for Q2--Q3, which aligns with the Lorentzian feature in the auto-PSD of Q3 shown in Fig.~\ref{fig:auto-PSD}, suggesting coupling to a shared TLF. The same feature appears in Q2--Q4 with reduced amplitude, and is no longer visible in Q2--Q5, showing a clear spatial decay of the correlation. The average correlation value for each pair is written in the upper left corner.
(b) Average correlation matrices for the frequency range $[1\times10^{-2}~\mathrm{Hz},\ 4~\text{Hz}]$, extracted under two different offset values $\Delta V_B$ of the interdot barrier voltage between Q2 and Q3 (voltage conditions different from those in (a)). In each cell, the central number is the maximum of the Bayesian posterior, with top and bottom numbers indicating 90\% confidence intervals. For positive $\Delta V_B$ (left), Q2 and Q3 are brought closer together, resulting in stronger correlations between them. For negative $\Delta V_B$ (right), Q2 and Q3 are pushed apart, their correlation weakens, while correlations Q1--Q2 and Q3--Q4 grow. The schematic (not to scale) qubit positions are illustrated above each matrix. The actual positional shifts can be estimated from the measured Zeeman energies and the known magnetic-field gradients (see Fig.~\ref{fig:device_f}). Overall, it is clear that noise correlations can be changed by gate voltages.}
\label{fig:correlation}
\end{figure}

We now focus on the spatial structure of noise correlations arising from charge noise at frequencies above $10^{-2}$~Hz, analyzing the correlation amplitudes after removing the global magnetic drift from the data. Unlike the global drift, these correlations are pair-dependent and decay with increasing qubit separation \cite{Donnelly2025}. In Fig.~\ref{fig:correlation_a}, we illustrate representative examples of the normalized cross-PSD for qubit pairs at varying distances: nearest neighbors (Q2--Q3), second neighbors (Q2--Q4), and third neighbors (Q2--Q5). For the nearest-neighbor pair, we observe a pronounced correlation peak reaching approximately 0.4 at $f \approx 0.2$~Hz. This peak coincides with the Lorentzian component in the auto-PSD of Q3 (Fig.~\ref{fig:auto-PSD}), suggesting shared coupling of qubits Q2 and Q3 to a single TLF, having a characteristic switching time of 0.7 seconds. The $\pi$ phase of the correlation ({Extended Data Fig.~\ref{exfig:full_matrix}}) shows that this TLF is located between Q2 and Q3: when its state switches, it induces qubit displacements whose projections along the array axis are opposite, resulting in opposite Zeeman-energy shifts \cite{Rojas-Arias2026}. The influence of this TLF persists in the Q2--Q4 pair, albeit with a reduced amplitude. For the more distant Q2--Q5 pair, the correlation peak becomes indistinct; however, the correlation phase ({Extended Data Fig.~\ref{exfig:full_matrix}}) remains near $\pi$, indicating detectable long-range correlation through the TLF's electric field.

Since the full spectral data for a qubit pair are rather complex, we introduce a simplified metric. With the details given in Methods, this metric represents data such as plotted in Fig.~\ref{fig:drift_corr} as a single number, representing the average level of correlation within a selected frequency range. This procedure allows us to construct \textit{average correlation} matrices, as shown in Fig.~\ref{fig:correlation_b} for the frequency range $[1\times10^{-2}~\mathrm{Hz},\ 4~\text{Hz}]$. These matrices reveal a clear pattern: the strongest correlations occur between nearest-neighbor qubits, and the correlation amplitude decreases systematically with distance. In contrast, the low-frequency correlation matrix for data without the drift removal, displayed in {Extended Data Fig.~\ref{exfig:matrix_drift}}, shows a global noise background, with correlations that do not decay with distance.

While correlations cannot be deterministically set, they can be tuned electrostatically through gate voltages, providing partial control over their magnitude. Specifically, the relative locations of neighboring qubits can be adjusted by changing the barrier voltage $\Delta V_B$ applied to the gate between them. In Fig.~\ref{fig:correlation_b}, we demonstrate this controlability by comparing two configurations with different $\Delta V_B$ values between Q2 and Q3. Leveraging knowledge of the local magnetic field gradient (Fig.~\ref{fig:device_f}; see Methods), we map the qubit energy shifts to relative qubit positions along the array. For a more positive $\Delta V_B$ (left panel), Q2 and Q3 move closer together (separation $\sim$108~nm), resulting in an average correlation of 0.57. A more negative $\Delta V_B$ (right panel) increases their separation to $\sim$137~nm, reducing the Q2--Q3 average correlation to 0.20. This displacement also alters correlations with adjacent qubits: Q2 approaches Q1 and Q3 approaches Q4, causing the Q1--Q2 correlation to rise from 0.33 to 0.55 and Q3--Q4 from 0.19 to 0.33. Intermediate $\Delta V_B$ values yield correlation amplitudes between these two limits (Extended Data Fig.~\ref{exfig:matrix_drift}), confirming their continuous control. 
Combined with the variability introduced by device-specific disorder configurations, gate tunability enables a statistical characterization of the distance dependence of noise correlations, which we explore next.

\section*{Scaling of correlations with qubit-qubit distance}

\begin{figure}
\centering
\includegraphics[scale=1]{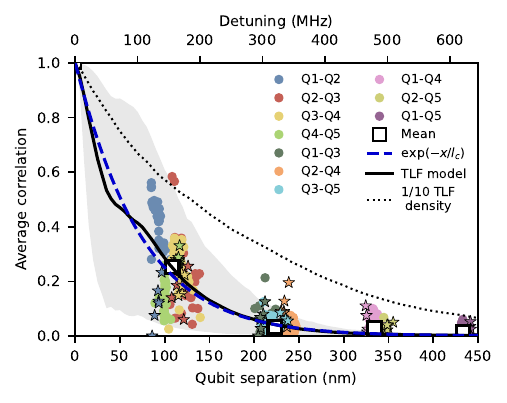}
\caption{\textbf{Scaling of noise correlations.}
Measured average correlations for all qubit pairs across multiple gate voltage configurations (colored symbols), evaluated after drift removal and over the frequency range $[1\times10^{-2}~\mathrm{Hz},\ 4~\text{Hz}]$, with circles and stars for the two cooldowns. White squares indicate the mean correlation for each neighbor order. 
The measured qubit energy detuning for each qubit pair is shown on the top horizontal axis and is converted to interqubit separation using the magnetic-field gradient along the qubit array axis of 0.051~mT/nm. The blue dashed line shows an exponential decay with correlation length $l_c=N_c L_q=81$~nm, corresponding to $N_c=0.75$ for an average nearest-neighbor qubit spacing of $L_q=108$~nm. The black curve shows the average correlation decay predicted by the TLF model with density $\rho_\text{TLF}=3\times10^{10}$~cm$^{-2}$, and the gray shaded region denotes the 10-90 percentile range across 1500 TLF ensembles, showing good agreement with the experimental data at short and intermediate separations. At larger interqubit distances, however, the measured correlations exhibit a statistically significant excess relative to the model, as analyzed in Extended Data Fig.~\ref{exfig:deviation}, indicating residual long-range correlations beyond the TLF-only description.
Our model has a single fit parameter, the density of TLFs. We show the sensitivity to this parameter by taking a density ten times lower than the best-fit value. The dotted line shows the predicted correlations, demonstrating that we can fit the TLF density with a resolution much better then an order-of-magnitude. 
}
\label{fig:scaling}
\end{figure}

{We have repeatedly measured noise correlations in our device over a span of eight months, during which the device underwent two cooldowns and was operated under varying gate-voltage conditions. In Fig.~\ref{fig:scaling}, we summarize the average correlation amplitude as a function of interqubit distance. During the long data-collection time, each qubit samples multiple noise environments due to changes in electrostatic configuration and the stochastic activity of nearby TLFs. While this variability (see Extended Data Fig.~\ref{exfig:auto-PSD_2nd} for an illustration of the changes upon a cooldown) leads to a spread in correlation values, a clear trend emerges: correlations decay with distance. This is well visible when averaging over qubit pairs according to their adjacency (white
squares in Fig.~\ref{fig:scaling}), and we have verified that these correlations are not induced by gate operations (see Extended Data Fig.~\ref{exfig:operated}). Despite strong variations, we have gathered a dataset large enough to allow for quantitative analysis.}

To model these observations, we performed Monte Carlo simulations of ensembles of dipolar TLFs that couple electrostatically to the qubits via screened Coulomb interactions (see Methods). The TLFs are assumed to be randomly distributed with a uniform two-dimensional density in the oxide layer underneath the metallic gates (Fig.~\ref{fig:device_e}). We compiled statistics over 1500 independent ensembles, each of which produces a correlation-versus-distance profile. The simulations reproduce the experimental trend, yielding a mean decay (black curve in Fig.~\ref{fig:scaling}) and a 10--90 percentile spread (gray shaded area) that capture the observed variability at short and intermediate separations. At the largest interqubit separations, a small residual correlation remains beyond the TLF prediction (see Extended Data Fig.~\ref{exfig:deviation}). The TLF density is the sole fit parameter in our model, and the extracted value of $3 \times 10^{10}$~cm$^{-2}$ agrees with previous reports \cite{Rojas-Arias2023,Culcer2013,Zimmermann1981}. The model predicts that correlations eventually follow a polynomial decay at large separations (Extended Data Fig.~\ref{exfig:model_log}) \cite{Rojas-Arias2023}. Within the experimentally accessible range, however, the data are well described by an exponential dependence $e^{-x/l_c}$ (blue dashed line in Fig.~\ref{fig:scaling}), from which we extract an effective correlation length of $l_c = 81$~nm. Having extracted the characteristic scale for spatial decay, we next examine the effects of such noise on quantum error correction. 

\section*{Impact of noise correlations on QEC performance}

The scaling of noise correlations with distance is critical for the feasibility of QEC \cite{Clemens2004}. We therefore evaluate QEC performance in the presence of noise with correlations as observed here. We analyze two codes: a repetition code that corrects phase-flip errors and a surface code. While the ultimate interest lies in the latter, we use the former as a model for which we can obtain analytical results. In addition, to isolate geometric effects, we compare one-dimensional and quasi-two-dimensional qubit layouts for the repetition code. Across these scenarios, we examine how the logical error rate scales with code distance under correlated and uncorrelated noise.

We first consider the $\llbracket N,1,N \rrbracket$ repetition code, consisting of $N$ physical qubits encoding one logical qubit with code distance $d=N$, allowing correction of up to $(d-1)/2$ qubit errors. Although the repetition code corrects only phase-flip errors, our focus here is specifically on spatially correlated dephasing. This is well motivated for spin qubits, where relaxation times $T_1$ can reach seconds~\cite{Koch2025,Camenzind2018}, while dephasing times $T_2^*$ are much shorter ($\approx 7~\mu\mathrm{s}$ in our device; see Extended Data Fig.~\ref{exfig:wide}).
To isolate the impact of correlated dephasing, we assume ideal syndrome extraction, decoding, and correction, so that no additional error channels are introduced during the QEC cycle. We adopt a physical phase-flip error rate $p=10^{-2}$, corresponding to dephasing accumulated during a cycle time $\tau = 1~\mu\mathrm{s}$ (see Methods).
For spatial correlations, we either use the correlation profile predicted by the TLF model (averaged over ensembles) or an exponential form
$c_{i,j} = \exp\left(-|\vec{r}_i-\vec{r}_j|/N_c L_q\right)$,
where $N_c$ is a characteristic decay length expressed in units of the qubit spacing $L_q$ (e.g., $N_c = 1$ corresponds to correlations that decay within nearest neighbors), and $\vec{r}_i$ is the position of qubit $i$.

\begin{figure}
\centering
\subfloat{\includegraphics[width=\textwidth]{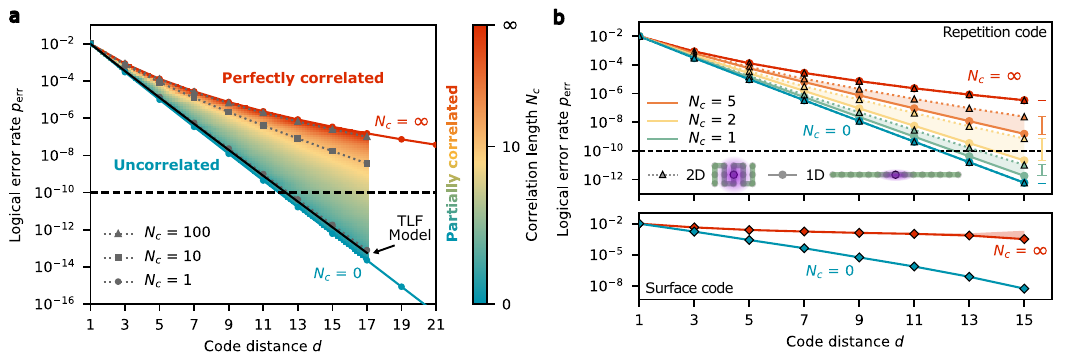}\label{fig:qec_a}}
\subfloat{\label{fig:qec_b}}
\caption{\textbf{QEC performance under correlated noise.}
(a) Logical error rate of the repetition code as a function of the code distance $d$ (equal to the number of physical qubits $N$ for this code) and correlation length $N_c$ for exponentially decaying correlations $c_{i,j} = \exp(-\left|\vec{r}_i-\vec{r}_j\right|/N_c L_q)$ (color gradient). Limiting cases of uncorrelated ($N_c \to 0$) and fully correlated ($N_c \to \infty$) noise are marked with colored circles; representative intermediate cases are shown with gray symbols. The solid black curve shows the logical error rate computed using the TLF model from Fig.~\ref{fig:scaling}. (b) Upper panel: Comparison of repetition-code performance for 1D and 2D qubit layouts of up to 15 qubits. Circles (triangles) show the logical error rate for the 1D (2D) array. The limiting cases of uncorrelated and perfectly correlated noise are identical for both architectures, but the 2D case exhibits notably worse performance for partially correlated noise. 
Lower panel: Surface-code logical error rate for perfectly correlated (red) and uncorrelated (blue) noise. Diamonds with solid lines denote the lower bound of the calculation described in Methods. The shaded region spans the interval between this lower bound and a corresponding upper bound, with the true logical error rate lying within the band. For uncorrelated noise the band is essentially invisible, indicating tight bounds. While uncorrelated noise yields exponential suppression of logical errors with increasing distance, perfectly correlated noise results in a reduction of less than two orders of magnitude when increasing the distance from $d=1$ to $d=15$. Because the number of data qubits scales as $N=d^2$ (reaching 225 qubits at $d=15$), this limited suppression under perfectly correlated noise comes at substantial qubit overhead. The qualitative agreement with the repetition-code results in the upper panel indicates that the impact of correlated noise is not specific to a particular QEC architecture.}\label{fig:qec}
\end{figure}

Figure~\ref{fig:qec_a} shows the logical error rate (see Methods) for different correlation regimes. For uncorrelated noise ($N_c \to 0$), the logical error rate decreases exponentially with code distance, consistent with standard QEC results~\cite{NielsenChuang,Aharonov1997,Fowler2012,Google2025}. In contrast, for perfectly correlated noise ($N_c \to \infty$), the suppression with distance is much weaker, demonstrating that global noise sources such as the magnetic field drift identified in this work can severely degrade QEC performance. Of particular relevance is the partially correlated regime dominated by charge noise. In this regime, the result obtained from the TLF-based model in Fig.~\ref{fig:scaling} is nearly indistinguishable from an exponential decay with $N_c \approx 1$, and only marginally worse than the uncorrelated limit. Within our model, logical error rates below $10^{-10}$, a commonly cited benchmark for fault-tolerant quantum computation~\cite{Beverland2022,Google2025}, can be achieved with a $d=13$ code, without additional overhead relative to the uncorrelated case. These estimates neglect relaxation processes ($T_1 = \infty$) and other hardware and decoding imperfections. In practice, such effects will reduce performance~\cite{Google2023,Google2025}, requiring larger code sizes where long-range correlations may become more relevant.

We now move beyond the repetition code implemented in a linear quantum dot array. While our ultimate goal is to assess surface-code architectures, we take an intermediate step. To retain the analytical tractability of the repetition code while capturing geometric effects relevant to two-dimensional layouts, we embed it in a planar qubit array topology \cite{Google2021}. The qubit arrangement is shown in the upper panel of Fig.~\ref{fig:qec_b}, which compares repetition codes embedded in 1D and 2D arrays. As expected, both layouts yield identical logical error rates in the limiting cases of uncorrelated and perfectly correlated noise. The array geometry becomes important for intermediate (the device size is the relevant scale) correlation lengths, where the 2D layout exhibits significantly higher logical error rates, due to smaller average qubit-qubit distances. Despite the more pronounced effects of correlations, the overhead remains modest in the regime relevant to our device. For a correlation length of $N_c=1$, reaching a logical error rate of $10^{-10}$ requires only a slight increase in code distance, from $d=13$ in 1D to $d=15$ in 2D. 

We finally look at the surface code. As already noted, here we can only derive estimates based on numerical sampling (see Methods), and only in the limiting cases of uncorrelated and perfectly correlated noise. Nevertheless, the results, plotted in the lower panel of Fig.~\ref{fig:qec_b}, closely resemble those of the repetition code: for uncorrelated noise, the logical error rate decreases rapidly with increasing code distance, whereas for perfectly correlated noise the suppression is significantly weaker and exhibits a pronounced flattening. An important difference is that the number of data qubits scales as $N=d^2$, so any reduction in logical-error suppression due to correlations translates into a larger qubit overhead. Although a full analysis of 2D QEC codes tailored to spin-qubit correlated noise is still needed \cite{Gravier2025,Gutierrez2025}, the qualitative agreement with the repetition-code results suggests that our simpler 2D model captures the essential geometric effects of spatial correlations. Since the TLF-based correlations observed in our device are far from the perfectly correlated limit, we conclude that the level of noise correlations typical for current semiconducting devices does not fundamentally limit surface-code implementations, even though these correlations are sizable.

\section*{Discussion}

Our results indicate that QEC remains viable under the level of correlated charge noise observed in our devices, a crucial finding for the scalability of spin qubits. Nevertheless, several considerations should be taken into account. First, while one might intuitively expect more compact qubit layouts to be advantageous, reduced spacing comes at the cost of stronger average correlations, which can significantly degrade QEC performance. Increasing the spacing between qubits therefore provides an effective route to reduce correlations. We have demonstrated that manipulation of the interqubit spacing can be implemented through gate-voltage tuning. Alternatively, it could also be done in sparse arrays by leaving empty quantum dots between active sites, using shuttling to enable two-qubit gates when needed \cite{Noiri2022a}.

Second, the impact of correlated noise must be considered jointly with the underlying physical error rate. Lower TLF densities reduce the physical error rate but are accompanied by longer correlation lengths (see the dotted line in Fig.~\ref{fig:scaling}), making correlated noise a more important consideration. Nevertheless, Extended Data Fig.~\ref{exfig:repetition_vs_density} shows that reducing the TLF density remains favorable overall: despite the increase in correlation length, the reduction in physical error rate leads to a faster decay of the logical error rate with code distance. This highlights the importance of evaluating correlated noise in conjunction with materials improvements, rather than treating correlation length alone as a figure of merit.

Third, although our analysis focused on low-frequency noise, we expect the conclusions to hold at higher frequencies. Charge noise in spin qubits exhibits similar spectral features up to the megahertz regime \cite{Connors2022,Yoneda2018,Eng2015} (see Extended Data Fig.~\ref{exfig:wide}), which is directly relevant on the timescale of individual QEC cycles with durations of order $1~\mu$s,  and there is no indication that these spatial correlations change systematically with frequency.
On the low-frequency end, the frequency range relevant to QEC is bounded by the frequency of device recalibrations, since recalibration effectively resets slow drifts. Practical quantum processors are expected to run for hours \cite{Gidney2021} (and even days \cite{Gidney2025}), but recalibration is time-consuming and therefore cannot be performed arbitrarily frequently. As a result, even recalibration intervals on the order of seconds set a low-frequency cutoff well within the range probed here, making the correlations characterized in this work directly relevant for QEC operation.

Fourth, global noise correlations arising from magnetic field drift impose a hard lower bound on the logical error rate ({Extended Data Fig.~\ref{exfig:limit_corr}}). Overcoming this limitation requires either engineering improvements that suppress magnetic drift below the target logical error rates, or alternative strategies such as encoding in decoherence-free subspaces (e.g., singlet-triplet or exchange-only qubits) or implementing passive or active feedback to cancel slow drifts. Although the impact of global magnetic noise is potentially severe, the availability of clear mitigation pathways suggests that it represents a technical challenge rather than a fundamental obstacle for the scalability of spin qubits.

The framework developed here provides a quantitative basis for characterizing correlated noise and assessing its impact on QEC. While our focus has been on Si/SiGe spin qubits, the correlation measurement protocol and QEC analysis are broadly applicable. Extending our approach to other qubit platforms will enable direct evaluation of how correlated noise limits their scalability toward fault-tolerant quantum computing.

\section*{Conclusions}

We have characterized spatial noise correlations in a five-qubit Si/SiGe spin-qubit array and established their implications for quantum error correction. Our measurements revealed two distinct regimes: global magnetic field drift that induces fully correlated energy fluctuations independent of distance, and charge noise arising from ensembles of two-level fluctuators that produces correlations decaying with separation and sensitive to the electrostatic environment. We introduced a microscopic model that explains the noise as due to an ensemble of two-level fluctuators. It reproduces the observed distance-dependence of the correlations and provides quantitative estimates of fluctuator density and correlation length.

Embedding these measured correlations into a repetition code allowed us to directly assess the impact of noise correlations on QEC. We find that global magnetic drift imposes a hard lower bound on the logical error rate, highlighting the need for improved hardware stability or mitigation strategies such as feedback or decoherence-free encoding. By contrast, the charge-noise correlations observed would have only minor effects on logical error rates, and stronger correlations can be mitigated by increased inter-qubit spacing.

Our results establish that correlated noise in silicon spin qubits, while present, does not constitute a fundamental barrier to fault-tolerant quantum computing at the levels observed here. This work sets quantitative benchmarks for correlated noise by relating measured spatial correlations to their impact on QEC. Extending this approach beyond silicon will be essential for evaluating how correlated noise constrains scalability in different architectures, an important step toward realizing fault-tolerant quantum processors.

\section*{Methods}

\subsection*{Device}\label{methods:device}

The device consists of gate-defined quantum dots formed in a 9~nm thick isotopically enriched ${}^{28}$Si quantum well with a residual ${}^{29}$Si concentration of 800~ppm. The quantum well lies beneath a 30~nm SiGe buffer and is capped by approximately 2~nm of native silicon oxide. A 15~nm Al$_2$O$_3$ dielectric layer was deposited by atomic layer deposition (ALD), followed by patterning of three overlapping aluminum gate layers using electron-beam lithography and lift-off. Each aluminum layer was oxidized to produce a thin native oxide that provides insulation between successive gate layers \cite{Wu2025}.

A micromagnet consisting of 5~nm of titanium and 250~nm of cobalt was fabricated on top of the gate stack, separated by a 30~nm ALD-grown Al$_2$O$_3$ layer. The micromagnet generates a static magnetic field gradient for qubit frequency selectivity and a slanting field that enables qubit operations using electric-dipole spin resonance (EDSR). The resulting gradient produces a Zeeman energy difference of $\Delta E_Z \approx 150$~MHz between adjacent qubits, in agreement with design expectations. The qubit frequencies vary nearly linearly across the array (Fig.~\ref{fig:device_f}), allowing their relative positions to be inferred from the measured Zeeman splittings. The slanting field provides Rabi frequencies up to 10~MHz, with optimal single-qubit fidelities typically achieved at drive frequencies around 3~MHz \cite{Wu2025}.

\subsection*{Qubit initialization, control, and readout}\label{methods:qubit_ctrl}
Qubit energies, dot confinements, and tunnel couplings are controlled via DC voltages applied using Qblox D5a digital-to-analog converters (DACs). Baseband pulses from Keysight M3201A arbitrary waveform generators (AWGs) pulse the qubits between operation and readout points. Qubit control pulses are generated by Keysight M3202A AWGs and mixed with local oscillator (LO) signals from a multi-channel signal generator (APMS20G4) using IQ mixers (2x MMIQ-0626HS, 1x MLIQ02181). Single-qubit gates are applied by delivering microwave bursts to the MW gate electrodes (Fig.~\ref{fig:device_d}).

Device operation follows a cycle of initialization, manipulation, and readout. Pauli spin blockade (PSB) is implemented by projection of Q2 (Q5) into qubit Q1 (Q4), with the charge state measured by adjacent charge  sensors (white ovals in Fig.~\ref{fig:device_d}). To enlarge the readout window, the array is tuned into the $(3,1,1,3,1)$ charge state. In this configuration, and due to mixing of the $T_0$ and singlet state, only spin parity is measured, not individual spin states \cite{Philips2022,Seedhouse2021}. Initialization of the Q1--Q2 (Q4--Q5) pairs is achieved by adiabatically separating the four-electron state from Q1 (Q4) into the empty Q2 (Q5), with the magnetic gradient ensuring preparation of the $\ket{\uparrow\downarrow}$ state. Q3 is initialized via a controlled rotation (CROT) gate using Q3 as control and freshly prepared Q2 as target. A subsequent PSB readout of the Q1--Q2 pair yields even or odd parity, from which the spin state of Q3 is inferred. A conditional $\pi$-rotation is then applied if the outcome is even, ensuring initialization of Q3 into the $\ket{\downarrow}$ state.

After initialization into $\ket{\uparrow\downarrow\downarrow\uparrow\downarrow}$, spin manipulations are performed. Because only spin parity is directly measurable, we use a two-stage cycle to reconstruct individual spin states. In the first stage, Q1, Q3, and Q5 are targeted, and Q1 (Q5) is read out via PSB projection with the initialized Q2 (Q4), providing an effective single-spin measurement. In the second stage, Q2, Q3, and Q4 are targeted, and readout is obtained by projecting Q2 (Q4) with respect to the initialized Q1 (Q5). Qubit Q3 is thus measured in both stages using the same non-demolition \cite{Yoneda2020,Nakajima2019} procedure applied for its initialization. This double measurement of Q3 improves statistics and is particularly important since the required CROT gate introduces an additional SPAM error channel.

During operation, the exchange interaction between qubits is typically limited to a few kHz \cite{Wu2025}, enabling independent and high-fidelity single-qubit gates. {For the measurements in Fig.~\ref{fig:correlation}, the exchange coupling is increased but remains well below the threshold for inducing conditional rotations, ensuring that single-qubit gates are unaffected by the spin state of neighboring qubits.}

\subsection*{Magnet field stability}
These experiments were performed using an Oxford Instruments CryoFree Superconducting Magnet operated in persistent mode. The solenoid is specified to have a field leakage rate of 1 part in $10^5$ per hour, corresponding to a qubit-energy drift of approximately 35~Hz/s at the experimental field of 0.45~T, assuming a $g$-factor of 2. In our measurements, we observe a lower drift rate, typically between 6 and 9~Hz/s. The drift rate differed between qubits and across measurements with varying electrostatic conditions. Notably, qubit frequencies return to their original values when persistent mode is disabled and the magnet is re-energized, indicating that the observed drift arises from slow changes in the external magnetic field rather than from time-dependent depolarization of the micromagnet or low-frequency charge noise. Outside of persistent mode, qubit measurements are not feasible due to a substantial increase in magnetic noise, caused by the PID controller of the magnet power supply.

To further investigate the origin of this drift, we performed measurements across multiple dilution refrigerators equipped with superconducting solenoid magnets. Across six dilution refrigerators from two independent manufacturers, global magnetic-field drifts were observed in four cases under comparable operating conditions, while two systems showed no measurable drift. In addition, magnets were exchanged between cryogenic systems. The qubit-energy reduction consistently followed the magnet rather than the cryostat, confirming that the effect originates from the solenoid itself. These observations indicate that the effect is reproducible for magnets that exhibit it and is not the result of an isolated failure. High-stability magnets with minimal leakage rates will be essential for the long experiments required in practical fault-tolerant quantum computing. For such applications, even smaller leakage rates, which are not currently detectable in our experiments, can become problematic since they define a lower bound of the logical error rate in QEC, as we show in Extended Data Fig.~\ref{exfig:limit_corr}.

Furthermore, the observed leakage rates remained stable over extended periods, enabling software-side correction of qubit energies in our experiments. In other measurements, including long-duration protocols such as Gate Set Tomography, we successfully applied linear qubit-energy drift corrections to account for this effect \cite{Wu2025}.

\subsection*{Auto- and cross-PSDs}
Our primary measure of noise is the two-sided power spectral density (PSD), defined as
\begin{align}
C_{\alpha,\beta}(f)=\int_{-\infty}^\infty \mathrm{d}t \braket{\delta\nu_\alpha(0)\delta\nu_\beta(t)}e^{2\pi i f t},
\label{eq:psd_def}
\end{align}
for the Zeeman energy fluctuations $\delta\nu_{\alpha,\beta}$ of qubits $\alpha$ and $\beta$, assumed to be stationary. For $\alpha=\beta$, $S_\alpha\equiv C_{\alpha,\alpha}$ is the auto-PSD, which describes how the noise power of a single qubit is distributed over frequency. For $\alpha\neq\beta$, it yields the cross-PSD, which captures correlations between the fluctuations of two qubits. The cross-PSD is generally complex, with magnitude and phase encoding the strength and character of the correlations.

To facilitate comparison across qubits with different noise strengths, we use the normalized cross-PSD,
\begin{align}
c_{\alpha,\beta}(f)\equiv\frac{C_{\alpha,\beta}(f)}{\sqrt{S_\alpha(f)S_\beta(f)}}.
\label{eq:cross_psd_def}
\end{align}
This dimensionless measure quantifies the fraction of noise shared between two qubits, ranging from 0 (uncorrelated) to 1 (perfectly correlated). The phase indicates whether the fluctuations are in-phase ($\mathrm{arg}(c_{\alpha,\beta})=0$), out-of-phase ($\mathrm{arg}(c_{\alpha,\beta})=\pi$), or exhibit a relative delay (other values).

\subsection*{Time-Resolved Qubit Noise Characterization}\label{methods:qubit_ctrl}
To characterize the noise acting on our device, we extract simultaneous qubit-energy time traces for all qubits using an interleaved Ramsey protocol (Extended Data Fig.~\ref{exfig:sequence}) adapted from Ref.~\cite{Rojas-Arias2023}. Each qubit is detuned by $\sim 2$~MHz to generate Ramsey fringes. The cycle consists of two operation stages, each comprising initialization, a Ramsey sequence ($\pi/2$ rotation, free evolution time $\tau_i$, and a final $\pi/2$ rotation), and projective readout. As described above, Q1, Q3, and Q5 are addressed sequentially during the first stage, while Q2, Q3, and Q4 are addressed during the second stage. Each cycle with evolution time $\tau_i$ thus yields two single-shot outcomes for Q3 and one for each of the other qubits.  

Records are obtained by repeating the interleaved Ramsey cycle while sweeping $\tau_i$ from 0 to $4~\mu$s in 100 steps. We then acquire long time series by repeatedly collecting records, with data sets ranging from 6 hours to over 24 hours. To manage memory constraints from the large readout volume, we average the raw time-resolved charge signal directly on the digitizer FPGA and store only a single averaged point per readout event.  

From each record, we extract two estimates of the qubit energy to correct for estimation errors \cite{Yoneda2023,Rojas-Arias2026}. An energy estimate can be written as $\nu^\text{est}=\nu^\text{true}+\nu^\text{err}$, where $\nu^\text{true}$ is the true qubit energy and $\nu^\text{err}$ the estimation error. The auto-PSD of the estimated energy then becomes $S_{\nu^\text{est}}(f) = S_{\nu^\text{true}}(f) + S_{\nu^\text{err}}(f)$, which overestimates the true PSD by $S_{\nu^\text{err}}(f)$. To remove this bias, we divide each record into subsets with even and odd evolution times $\tau_i$, obtaining two independent estimates $\nu_\alpha^{(e)}$ and $\nu_\alpha^{(o)}$ for each qubit $\alpha$ via Bayesian estimation of qubit energies \cite{Delbecq2016,Nakajima2020}. From these we define the average estimate $\bar{\nu}_\alpha = {(\nu_\alpha^{(o)}+\nu_\alpha^{(e)})}/{2}$, and the estimation error $\nu_\alpha^\text{err} = {(\nu_\alpha^{(o)}-\nu_\alpha^{(e)})}/{2}$. Repeated acquisition yields time traces for both $\bar{\nu}_\alpha$ and $\nu_\alpha^\text{err}$, from which we compute auto-PSDs using the Bayesian estimation of correlation functions from Ref.~\cite{Gutierrez-Rubio2022}. The corrected auto-PSD for each qubit is then given by $S_{\bar{\nu}}(f) - S_{\nu^\text{err}}(f)$ calculated also via Bayesian estimation. Error bars are derived from the Bayesian distributions at 90\% confidence.  

Cross-PSDs are not affected by estimation errors. However, normalization introduces a dependence on the auto-PSDs in the denominator of Eq.~\eqref{eq:cross_psd_def}, and therefore the normalized cross-PSDs are computed using the corrected auto-PSDs as input. Confidence intervals are obtained by propagating the error bars of the contributing auto- and cross-PSDs.  

\subsection*{Auto-PSD of magnetic drift}

We model slow magnetic-field drift as a deterministic linear trend in the qubit frequency, $\delta\nu(t)=\alpha t$, measured over a finite acquisition time $T$. In our experiment $\delta\nu$ corresponds to the qubit Zeeman energy, with a typical drift rate $\alpha \sim 8~\mathrm{Hz/s}$. Such a process is explicitly non-stationary, so the Wiener-Khinchin definition of the PSD (Eq.~\eqref{eq:psd_def}) does not apply.

Instead, we characterize the spectral weight of the drift by computing the squared magnitude of the finite-time Fourier transform,
\begin{align}
S(f)=\frac{|\delta\tilde{\nu}(f)|^2}{T},
\end{align}
where
\begin{align}
\delta\tilde{\nu}(f)\equiv\int_0^T \mathrm{d}t\ \delta\nu(t)\ e^{2\pi i f t}.
\end{align}
For a linear drift this yields
\begin{align}
S(f)=\frac{\alpha^2}{16\pi^4 f^4 T}\Big[2+4\pi^2 f^2 T^2-2\cos(2\pi f T)-4\pi f T\sin(2\pi f T)\Big].
\end{align}
At frequencies large compared to the inverse acquisition time, $f \gg (2\pi T)^{-1}$, this expression simplifies to
\begin{align}
S(f)\approx\frac{\alpha^2 T}{4\pi^2 f^2},
\label{eq:drift_psd}
\end{align}
corresponding to an effective $1/f^2$ spectrum. In our analysis this is the relevant regime: the auto-PSD is evaluated only at frequencies $f \geq 1/T$, while lower frequencies are intrinsically unresolved due to the finite measurement duration. Equation~\eqref{eq:drift_psd} therefore provides the appropriate reference for the low-frequency magnetic drift contribution and is plotted as dashed lines in Fig.~\ref{fig:auto-PSD}.

\subsection*{Average correlation}

We now extend the framework of Ref.~\cite{Gutierrez-Rubio2022} to define a metric that condenses frequency-resolved correlation information into a single quantity for a given qubit pair. For completeness and to establish notation, we first restate the definitions and results from Ref.~\cite{Gutierrez-Rubio2022} that are used to compute probability distributions for cross-PSDs from measured data, such as those shown in Extended Data Fig.~\ref{exfig:full_matrix}. In our case, the input data consist of time traces of qubit energies. Each trace is divided into $M$ batches of length $N$, denoted $\nu_\alpha(j,m)$, where $m=0,\ldots,M-1$ indexes the batch and $j=0,\ldots,N-1$ labels the discrete time points $t_j = j\Delta t$, with $\Delta t$ the sampling interval.

The discrete Fourier transform of each batch is defined as  
\begin{align}
\lambda_\alpha(k,m)=\frac{1}{\sqrt{N}}\sum_{j=0}^{N-1}\nu_\alpha(j,m)\exp\left(-i\frac{2\pi j k}{N}\right),
\end{align}
with $k=0,\ldots,N-1$ corresponding to frequencies $f_k=k/(N\Delta t)$.  

The Bayesian posterior for the normalized cross-PSD (magnitude $s_k$ and phase $\phi_k$) given the data is, Eq.~(40) of Ref.~\cite{Gutierrez-Rubio2022}, for $k\neq 0$, 
\begin{align}
P(s_k,\phi_k|\mathrm{data})\propto (1-s_k^2)^M(1-q_{s_k\phi_k})^{\frac{1}{2}-2M} 
F\!\left(\begin{array}{c}
\frac{1}{2},\frac{1}{2} \\ 
2M+\frac{1}{2}
\end{array}\Bigg|\frac{1+q_{s_k\phi_k}}{2}\right),
\label{eq:bayesian}
\end{align}
where $F$ is the hypergeometric function, and $q_{s_k\phi_k}\equiv s_k\bar{s}_k\cos(\phi_k-\bar{\phi}_k)$. The sufficient statistics $\bar{s}_k$ and $\bar{\phi}_k$ are computed as  
\begin{align}
\bar{s}_k=\frac{\bar{\Lambda}^{\alpha\beta}_k}{\sqrt{\bar{\Lambda}_k^\alpha\bar{\Lambda}_k^\beta}},\quad
\bar{\phi}_k=\arg\!\left(\bar{\Lambda}_k^{\alpha\beta}\right),
\label{eq:sufficient_stats}
\end{align}
with
\begin{align}
\bar{\Lambda}_k^{\alpha\beta}=(1/M)\sum_{m=0}^{M-1}\lambda_\alpha(k,m)\lambda_\beta^*(k,m)
\end{align}
and $\bar{\Lambda}_k^\alpha=\bar{\Lambda}_k^{\alpha\alpha}$. These $\bar{\Lambda}$ correspond to the standard PSD estimators, which serve as input for the Bayesian procedure to yield a probability distribution and confidence intervals. Marginalizing Eq.~\eqref{eq:bayesian} over $s_k$ (or $\phi_k$) provides the distribution for the phase (or magnitude) alone.  

We now introduce the key extension beyond Ref.~\cite{Gutierrez-Rubio2022} by condensing frequency-resolved correlation information into a single effective parameter. Specifically, we define an \textit{average correlation} over a finite frequency range $k_0 \leq k \leq k_1$, which enables the construction of correlation matrices (Fig.~\ref{fig:correlation}) and the analysis of distance-dependent trends
(Fig.~\ref{fig:scaling}). The procedure parallels the single-frequency case but replaces $\bar{\Lambda}_k$ by effective statistics that combine both batches $m$ and frequency bins $k$:
\begin{align}
\bar{\Lambda}_\text{eff}^{\alpha\beta}=\frac{1}{K}\sum_{k=k_0}^{k_1}\bar{\Lambda}_k^{\alpha\beta},
\label{eq:lambda_eff}
\end{align}
where $K$ is the number of frequency bins in the interval. Effective quantities $\bar{s}_\text{eff}$ and $\bar{\phi}_\text{eff}$ are then obtained via Eq.~\eqref{eq:sufficient_stats} using $\bar{\Lambda}_\text{eff}$ in place of $\bar{\Lambda}_k$, with the auto-PSDs in the denominator corrected for estimation errors as described above.

Finally, when evaluating Eq.~\eqref{eq:bayesian} for average correlations, the number of batches $M$ must be adjusted to account for the increased amount of data obtained by combining multiple frequency bins. This adjustment is not equivalent to simply setting $M_\text{eff} = MK$, since the effective statistical weight depends on how the spectral weight is distributed across frequencies. To properly capture this, we rescale $M$ using the inverse participation ratio,
\begin{align}
M_\text{eff}=M\frac{\left(\sum_{k=k_0}^{k_1}|\bar{\Lambda}_k^{\alpha\beta}|\right)^2}{\sum_{k=k_0}^{k_1}|\bar{\Lambda}_k^{\alpha\beta}|^2}.
\label{eq:m_eff}
\end{align}
Evaluating Eq.~\eqref{eq:m_eff} in two limiting cases shows that the inverse participation ratio provides an appropriate definition of the effective number of frequency batches. First, consider a white spectrum for which $\Lambda_k^{\alpha\beta} = \Lambda$ for all $k$. In this case, all frequencies contribute equally and $M_\text{eff} = MK$. In the opposite limit, where only a single frequency component contributes, $\Lambda_k^{\alpha\beta} = \Lambda_{k'}\delta_{k,k'}$, one finds $M_\text{eff} = M$. This correctly reflects that no additional statistical weight is gained by combining frequencies, and only the batches used to estimate $\Lambda_{k'}$ contribute. These limits illustrate that $M_\text{eff}$ faithfully captures the effective statistical weight associated with combining multiple frequency components.

\subsection*{Two-level fluctuator (TLF) model}

We model charge fluctuators as dipoles located in the oxide layer of the device (Fig.~\ref{fig:device_e}). Each TLF $i$ switches between two charge states $q_i=\pm 1$ with characteristic switching time $t_c^{(i)}$. The electric field at the position $\vec{r}_\alpha$ of qubit $\alpha$ due to a dipole at $\vec{r}_i$ with moment $\vec{p}_i$ is given by  
\begin{align}
\vec{E}_i(t)=\frac{1}{4\pi\epsilon\epsilon_0}\left[\frac{3\big(\vec{p}_i\cdot(\vec{r}_\alpha-\vec{r}_i)\big)(\vec{r}_\alpha-\vec{r}_i)}{|\vec{r}_\alpha-\vec{r}_i|^5}-\frac{\vec{p}_i}{|\vec{r}_\alpha-\vec{r}_i|^3}\right]q_i(t),
\end{align}
where $\epsilon$ is the effective dielectric constant of the multilayer and $\epsilon_0$ is the vacuum permittivity. Screening from the metallic gates is included by introducing an image dipole $\vec{p}_i^\text{im}=-p_i^x\hat{x}-p_i^y\hat{y}+p_i^z\hat{z}$ located at $\vec{r}_i^\text{im}=x_i\hat{x}+y_i\hat{y}+(2h_z-z_i)\hat{z}$, with $h_z=51$~nm the vertical distance between the Si quantum well and the gates. The total field at the qubit is $\vec{E}_i+\vec{E}_i^\text{im}$.  

Switching of the TLFs induces electric-field jumps that displace the electron wavefunction in the dot. For qubit $\alpha$, the displacement is  
\begin{align}
\delta \vec{r}_\alpha(t)=\frac{e}{m\omega_0^2}\sum_{i}\big(E_i^x(t)\hat{x}+ E_i^y(t)\hat{y}\big),
\end{align}
where $m=0.19m_0$ is the effective electron mass in Si, $m_0$ the free-electron mass, and $\hbar\omega_0$ the in-plane orbital confinement energy, assumed isotropic. The out-of-plane displacement is neglected due to strong confinement from the quantum well.  

These displacements modulate the Zeeman splitting through the micromagnet field gradient, producing qubit-energy fluctuations
\begin{align}
\delta\nu_\alpha(t)=\frac{g\mu_B}{h}\nabla B_y\cdot\delta\vec{r}_\alpha(t),
\label{eq:delta_nu}
\end{align}
with $g=2$ the electron $g$-factor in Si, $\mu_B$ the Bohr magneton, $h$ Planck's constant, and $\nabla B_y=(\partial B_y/\partial x,\ \partial B_y/\partial y)=(0.051,\ 0.001)$~mT/nm from device simulations. The $y$-component is aligned with the external in-plane magnetic field defining the quantization axis, and $x$ is the direction along the qubit array.  

We treat TLFs as independent, with correlation function $\langle q_i(t')q_j(t'+t)\rangle=\delta_{ij}\exp\left(-|t|/t_c^{(i)}\right)$. Using this relation, auto- and cross-PSDs are calculated according to Eq.~\eqref{eq:psd_def}. TLF ensembles are generated randomly in a simulation volume $V=L_xL_yL_z$ with $L_x=L_y=4~\mu$m and $L_z=17$~nm (oxide thickness), with uniform density. Switching times are sampled from a log-uniform distribution, which ensures equal statistical weight per decade of $t_c^{(i)}$ and captures the broad range of fluctuation rates typically observed in $1/f$ charge noise \cite{Paladino2014}. Dipole orientations are random, with magnitude $|\vec{p}|=ed_\text{dip}$ and dipole size $d_\text{dip}=0.17$~nm (Ref.~\cite{Hung2022} reports the projection $\braket{|p^z|}=0.96~e$\AA, and we take $|\vec{p}|=\sqrt{3}\braket{|p^z|}$). For each set of parameters, we generate 1500 ensembles and compute average auto- and cross-PSDs for qubit pairs at different separations.  

The TLF density $\rho_\text{TLF}$ is treated as a fitting parameter for the correlation decay (Fig.~\ref{fig:scaling}), yielding $\rho_\text{TLF}=3\times 10^{10}$~cm$^{-2}$. The confinement energy $\hbar\omega_0$ is adjusted such that the average auto-PSD integrated in the frequency range $[10^{-2}$~Hz$,4$~Hz$]$ matches the measured value, giving $\hbar\omega_0=1.2$~meV, consistent with previous reports \cite{Camenzind2019} and device simulations.

The dipolar TLF model not only reproduces the exponential-like decay of correlations observed in Fig.~\ref{fig:scaling}, but also predicts a transition to a polynomial dependence at large interqubit separations, with an asymptotic behavior $\propto x^{-4.2}$ (Extended Data Fig.~\ref{exfig:model_log}) \cite{Rojas-Arias2023}. In the absence of screening, correlations decay much more slowly, following a $\propto x^{-3}$ law characteristic of unscreened dipoles. Screening from the metallic gates therefore suppresses the long-range component of correlated charge noise, a factor that is critical when assessing scalability at the QEC level (Fig.~\ref{fig:qec}) \cite{Rojas-Arias2023}.

We also fitted the correlation decay in Fig.~\ref{fig:scaling} using the screened charge-trap model that we previously reported in Ref.~\cite{Rojas-Arias2023}. This model yields a TLF density of $\rho_\text{TLF}=5\times10^{10}$~cm$^{-2}$, comparable to the value extracted from the dipolar model. However, reproducing the measured auto-PSD magnitude within this framework requires confinement energies $\hbar\omega_0>10$~meV, which are incompatible with our device parameters. We therefore rule out the charge-trap scenario and conclude that the relevant fluctuators are screened charge dipoles, consistent with recent suggestions \cite{Ye2024}.

This comparison illustrates that cross-PSD analysis can discriminate between competing microscopic models of charge noise, as we suggested previously in Ref.~\cite{Rojas-Arias2026}. The agreement in extracted TLF densities across models indicates that $\rho_\text{TLF}$ is a robust metric for characterizing charge noise across devices, providing a quantitative benchmark of oxide quality.

\subsection*{Repetition code error rate}

We evaluate the performance of an $N$-qubit repetition code that corrects phase-flip errors. In this analysis, we focus exclusively on correlated dephasing acting on the data qubits. We neglect correlations between data and ancillary qubits used for syndrome extraction and assume ideal syndrome extraction, decoding, and correction. Thus, the logical error rate reflects solely the impact of spatially correlated dephasing during the QEC cycle. Logical states are encoded as product states,
\begin{align}
\ket{0_L}=\prod_{k=1}^N\ket{+}_k,\qquad \ket{1_L}=\prod_{k=1}^N\ket{-}_k,
\end{align}
so that the code is of type $\llbracket d,1,d \rrbracket$, with $d=N$ the code distance, and can correct up to $(N-1)/2$ phase-flip errors.  During free evolution, qubit $k$ accumulates a stochastic phase
\begin{align}
\phi_k=\int_0^\tau \text{d}t\ \delta\nu_k(t),
\end{align}
arising from fluctuations in its Zeeman energy $\delta\nu_k(t)$. The corresponding dephasing operator for the whole qubit array is
\begin{align}
D=\bigotimes_{k=1}^N\big(\alpha_k+i\beta_k\sigma_k^z\big),
\end{align}
with $\alpha_k=\cos(\phi_k/2)$ and $\beta_k=\sin(\phi_k/2)$. Expanding $D$ generates terms corresponding to errors on zero, one, two, ..., up to $N$ qubits. Errors affecting at most $(N-1)/2$ qubits are correctable; higher-order errors lead to logical error. Our goal is to calculate the probability $p_n$ of an error involving $n$ qubits.  

Let $K_n$ denote the set of all $n$-element subsets of the qubit array indexes $[N]=(1, 2, ... N)$. Thus, an element of this set, $K_n=\{\mathcal{S}\subseteq [N]:|\mathcal{S}|=n\}$ is a specific choice of $n$ qubits from the $N$ qubit array. Below, it will represent the qubits that experience an error. The number of such configurations is given by the binomial coefficient $|K_n|={N\choose n}$. The probability of an $n$-qubit error is then
\begin{align}
p_n=\sum_{\mathcal{S}\in K_n}
\Big\langle \prod_{i\in\mathcal{S}}|\beta_i|^2 \prod_{j\notin\mathcal{S}}|\alpha_j|^2 \Big\rangle,
\label{eq:pn_def}
\end{align}
where $\langle\cdots\rangle$ denotes an average over noise realizations of a specific TLF ensemble. Using $|\alpha_k|^2=1-|\beta_k|^2$ and $|\beta_k|^2=(1-\cos\phi_k)/2$, these averages reduce to correlators of products of $\cos\phi_k$.  

To evaluate Eq.~\eqref{eq:pn_def}, we require averages of products of cosine factors. For any index set $\mathcal{U}\subseteq[N]$ with $|\mathcal{U}|=m$,
\begin{align}
\Big\langle \prod_{j\in\mathcal{U}}\cos\phi_j \Big\rangle
=\Big\langle \prod_{j\in\mathcal{U}}\tfrac{1}{2}\big(e^{i\phi_j}+e^{-i\phi_j}\big)\Big\rangle.
\end{align}
Expanding yields $2^m$ terms, each corresponding to a choice of sign for every factor $e^{\pm i\phi_j}$. We denote these sign assignments by vectors $\mathbf{s}=(s_1,\ldots,s_m)\in\{\pm1\}^m$, with $\{\pm1\}^m$ the set of all $m$-component vectors with entries $s_j=\pm1$. The average can then be written compactly as
\begin{align}
\Big\langle \prod_{j\in\mathcal{U}}\cos\phi_j \Big\rangle
=\frac{1}{2^m}\sum_{\mathbf{s}\in\{\pm1\}^m}
\Big\langle \exp\!\Big(i\sum_{j\in\mathcal{U}} s_j\phi_j\Big)\Big\rangle.
\label{eq:cos_prod}
\end{align}

Each term in Eq.~\eqref{eq:cos_prod} is the expectation of an exponential of a linear combination of phases. Assuming the phases $\{\phi_k\}$ are jointly Gaussian with zero mean, we have for any $X=\sum_j a_j\phi_j$, $\langle e^{iX}\rangle=\exp\!\big(-\tfrac{1}{2}\langle X^2\rangle\big)$ \cite{Kubo1962}. Thus,
\begin{align}
\Big\langle \exp\!\Big(i\sum_{j\in\mathcal{U}} s_j\phi_j\Big)\Big\rangle
=\exp\!\left[-\tfrac{1}{2}\Big\langle\Big(\sum_{j\in\mathcal{U}} s_j\phi_j\Big)^2\Big\rangle\right].
\label{eq:gauss_char}
\end{align}

To make the sign structure explicit, we partition $\mathcal{U}$ into disjoint subsets $\mathcal{Q}_+=\{j\in\mathcal{U}:s_j=+1\}$ and $\mathcal{Q}_-=\{j\in\mathcal{U}:s_j=-1\}$. Then
\begin{align}
\Big\langle e^{i\sum_{j\in\mathcal{Q}_+}\phi_j} e^{-i\sum_{j\in\mathcal{Q}_-}\phi_j}\Big\rangle
=\exp\!\left[-\tfrac{1}{2}\Big\langle\Big(\sum_{j\in\mathcal{Q}_+}\phi_j-\sum_{j\in\mathcal{Q}_-}\phi_j\Big)^2\Big\rangle\right].
\label{eq:QplusQminus}
\end{align}
Equation~\eqref{eq:QplusQminus} depends only on two-point correlators of the accumulated phases, $\langle \phi_i \phi_j \rangle$, which in turn are determined by the specific realization of the two-level fluctuator (TLF) environment. In principle, one could obtain the logical error rate by first evaluating $p_n$ for a given TLF realization and then averaging over TLF ensembles. However, because Eq.~\eqref{eq:pn_def} is nonlinear in the phase correlators, this ensemble average is not analytically tractable.

Instead, we characterize the typical behavior by replacing the phase correlators in Eq.~\eqref{eq:QplusQminus} with their ensemble-averaged values, denoted by $\overline{(\cdot)}^{\mathrm{TLF}}$. This procedure yields an effective description of the average error statistics induced by the fluctuator ensemble. We assume that all qubits experience identical noise, such that
$\overline{\langle \phi_j^2 \rangle}^{\mathrm{TLF}} = \overline{\langle \phi^2 \rangle}^{\mathrm{TLF}}$,
corresponding to a common ensemble-averaged auto-PSD $\overline{S(f)}^{\mathrm{TLF}}$. We further assume that the normalized cross-PSDs are frequency independent, as expected when averaging over many independent TLF ensembles \cite{Rojas-Arias2023}. Under these assumptions, the ensemble-averaged phase correlators are
\begin{align}
\overline{\langle \phi^2 \rangle}^{\mathrm{TLF}} &=
4 \int_{-\infty}^{\infty} \mathrm{d}f\ 
\overline{S(f)}^{\mathrm{TLF}}
\frac{\sin^2(\pi f \tau)}{f^2},
\label{eq:auto_identical}\\
\overline{\langle \phi_i \phi_j \rangle}^{\mathrm{TLF}} &=
4 \int_{-\infty}^{\infty} \mathrm{d}f\ 
\overline{C_{i,j}(f)}^{\mathrm{TLF}}
\frac{\sin^2(\pi f \tau)}{f^2}
= c_{i,j}\ \overline{\langle \phi^2 \rangle}^{\mathrm{TLF}}.
\label{eq:cross_identical}
\end{align}

Because the correlator $\langle \phi_i \phi_j \rangle$ can take either sign depending on the relative location of the qubits and the fluctuators, its TLF ensemble average may be strongly suppressed by cancellations between positive and negative contributions. In such cases, the ensemble average does not reflect the typical magnitude of correlations present in a given device, analogous to a fluctuating quantity with zero mean but finite variance. To obtain a more representative estimate of the error statistics relevant for QEC, we therefore use the ensemble average of the absolute value, $\overline{|\langle \phi_i \phi_j \rangle|}^{\mathrm{TLF}}$, as the effective input to Eq.~\eqref{eq:QplusQminus}.

With these inputs, the probabilities $p_n$ follow from Eq.~\eqref{eq:pn_def} using the Gaussian averages above. The calculation depends on two ingredients only: the single-qubit phase statistics and the spatial correlation profile encoded in the coefficients $c_{i,j}$, which we assume to depend solely on the interqubit separation $|\vec r_i - \vec r_j|$. The single-qubit contribution is conveniently parametrized by the phase-flip probability
\begin{align}
p = \tfrac{1}{2}\left(1 - e^{-\langle \phi^2 \rangle / 2}\right)= \tfrac{1}{2}\left(1 - e^{-\tau^2/T_2^{*2}}\right),
\label{eq:single_error}
\end{align}
which we use as the practical input parameter in place of $\langle \phi^2 \rangle$.

Finally, the logical error probability of the repetition code is obtained by summing over all uncorrectable error events,
\begin{align}
p_{\mathrm{err}} =
\sum_{n=\frac{N+1}{2}}^{N} p_n
= 1 - \sum_{n=0}^{\frac{N-1}{2}} p_n .
\label{eq:repetition_error}
\end{align}

In practice, the inputs to the repetition-code calculation are extracted directly from the measured auto- and cross-PSDs of the qubit energy fluctuations. The auto-PSDs determine the single-qubit phase variance and hence the physical error rate $p$ via Eq.~\eqref{eq:single_error}. We consider a QEC cycle time $\tau=1~\mu$s \cite{Google2025}; using an average coherence time $T_2^*\approx7~\mu$s (for an integration time of $\sim100$~s, see Extended Data Fig.~\ref{exfig:wide}), this corresponds to a physical error rate $p\approx10^{-2}$. The spatial correlation profile $c_{i,j}$ is taken either as exponentially decaying with interqubit separation or as the average correlation extracted from the TLF model that reproduces the measured correlation decay in Fig.~\ref{fig:scaling}. With these inputs specified, the logical error rate follows from Eq.~\eqref{eq:repetition_error}, with the $n$-qubit error probabilities $p_n$ evaluated using Eqs.~\eqref{eq:pn_def}--\eqref{eq:QplusQminus}.

\subsection*{Surface code error rate}

The evaluation of the logical error rate for the surface code follows the same general strategy as for the repetition code, but with important differences. First, the number of data qubits scales quadratically with the code distance, yielding a $\llbracket d^2,1,d \rrbracket$ code with $N=d^2$ physical qubits. Second, unlike the repetition code, the surface code can correct a subset of error configurations with order exceeding $(d-1)/2$. Third, which high-order errors are correctable depends explicitly on the decoder.

Taking these considerations into account, the logical error probability of a distance-$d$ surface code for a given decoder can be written as
\begin{align}
p_\text{err}=\sum_{n=\frac{d+1}{2}}^{N} \sum_{i=1}^{\binom{N}{n}} \gamma_{n}^{(i)} p_{n}^{(i)},
\label{eq:qec_expansion}
\end{align}
where the index $i$ labels all configurations of $n$ errors among $N$ qubits, $p_n^{(i)}$ is the probability of a specific configuration, and $\gamma_n^{(i)}$ is a binary indicator equal to 1 if that configuration leads to a logical error under the chosen decoder and 0 otherwise. For the repetition code, all errors with $n \ge (d+1)/2$ are uncorrectable, so $\gamma_n^{(i)}=1$ for all such configurations, reducing Eq.~\eqref{eq:qec_expansion} to Eq.~\eqref{eq:repetition_error}. For the surface code, by contrast, determining $\gamma_n^{(i)}$ is nontrivial, and the total number of configurations $\binom{N}{n}$ grows prohibitively with $d$, making an exact evaluation intractable.

We therefore restrict our analysis to two limiting noise regimes in which Eq.~\eqref{eq:qec_expansion} simplifies: uncorrelated noise and perfectly correlated noise. In both cases, all error configurations of a given order $n$ are equally likely, so that $p_n^{(i)}$ is independent of $i$. This allows us to write
\begin{align}
p_\text{err}=\sum_{n=\frac{d+1}{2}}^{N} F_n p_n,
\label{eq:surface_expansion}
\end{align}
where
\begin{align}
F_n=\frac{1}{{N\choose n}}\sum_{i=1}^{\binom{N}{n}} \gamma_n^{(i)}
\end{align}
is the fraction of uncorrectable error configurations of order $n$ for the chosen decoder.

The probability $p_n$ of an $n$-qubit error is given by Eq.~\eqref{eq:pn_def}. For uncorrelated noise this reduces to the binomial distribution
\begin{align}
p_n=\binom{N}{n} p^n (1-p)^{N-n},
\end{align}
while for perfectly correlated noise it takes the form
\begin{subequations}
\begin{align}
p_n&=\binom{N}{n} \sum_{k=0}^{N-n} \binom{N-n}{k}\frac{1}{4^{n+k}}\sum_{l=0}^{2(n+k)}\binom{2(n+k)}{l}(-1)^{n-l}(1-2p)^{(n+k-l)^2}\\
  &\approx\binom{N}{n} p^n \sum_{k=0}^{N-n} \binom{N-n}{k} [2(n+k)-1]!!(-p)^k,
\end{align}
\end{subequations}
with $p$ the physical phase-flip error probability defined in Eq.~\eqref{eq:single_error}, and the last approximation valid for $p\ll1$.

To estimate $F_n$, we use Monte Carlo sampling. For each order $n$, we randomly generate $n_\text{samples}=15000$ phase-flip error configurations and evaluate whether they result in a logical error for a distance-$d$ surface code. Syndrome extraction is simulated using \texttt{Stim} \cite{Gidney2021a}, and decoding is performed using minimum-weight perfect matching implemented via \texttt{pyMatching} \cite{Higgott2025,Higgott2021}. The fraction of sampled configurations leading to a logical error provides an estimate of $F_n$.

Because the number of terms in Eq.~\eqref{eq:surface_expansion} grows rapidly with $d$, we employ a subset-sampling strategy \cite{Heussen2024,Gutierrez2019} in which only the lowest-order contributions are evaluated explicitly. Introducing a cutoff $n_\text{max}$, we define a truncated expansion that yields a lower bound on the logical error rate,
\begin{align}
p_\text{err}^\text{LB}=\sum_{n=\frac{d+1}{2}}^{\frac{d+1}{2}+n_\text{max}} F_n p_n.
\end{align}
An upper bound is obtained by assigning unit weight to all neglected higher-order terms,
\begin{align}
p_\text{err}^\text{UB}=p_\text{err}^\text{LB}+\sum_{n=\frac{d+3}{2}+n_\text{max}}^{N} p_n.
\end{align}
The true logical error rate lies between these two bounds.

We use $n_\text{max}=14$ throughout, which yields narrow bounds for the distances considered, providing a good approximation to the logical error rate. The resulting bounds are shown in the lower panel of Fig.~\ref{fig:qec_b}. As in the repetition code, we find a strong difference between uncorrelated and perfectly correlated noise, with the latter severely suppressing the benefits of increasing code distance.

\backmatter

\bmhead{Acknowledgements}

We thank Mauricio Guti\'errez for helpful discussions on quantum error correction. We acknowledge R.~Kuroda for support with sample fabrication. This work was supported financially by the Japan Science and Technology Agency (JST) Moonshot R\&D Program (Grant No. JPMJMS226B), Core Research for Evolutional Science and Technology (CREST) (Grant No. JPMJCR1675), MEXT Quantum Leap Flagship Program (MEXT Q-LEAP) (Grant No. JPMXS0118069228), and JSPS KAKENHI (Grant Nos. 18H01819, 20H00237, and 23K26483). This work was also supported by the Swiss National Science Foundation and NCCR SPIN (Grant No. 225153). T.N. acknowledges support from JST PRESTO (Grant No. JPMJPR2017). A.N. acknowledges support from JST PRESTO (Grant No. JPMJPR23F8). Y.H.W. acknowledges support from RIKEN's IPA program. D.L. acknowledges support from the Deanship of Research (Grant No. CUP25102) and the Quantum Center (Grant No. INQC2500).

\section*{Declarations}
\subsection*{Conflict of interest}
The authors declare no competing interests.
\subsection*{Author Contribution}
J.S.R.-A, L.C.C., Y.H.W., and P.S. conceived the project. Y.H.W. and L.C.C. performed the experiments. A.N. fabricated the device. J.S.R.-A., Y.H.W., and L.C.C. analyzed the data. J.S.R.-A. and P.S. developed the theoretical models. K.T., A.N., T.K., and T.N. contributed to data acquisition and discussions. G.S. developed and supplied the isotopically enriched $^{28}$Si/SiGe heterostructure. J.S.R.-A. and L.C.C. wrote the manuscript with input from all authors. D.L. and S.T. supervised the project.

\begin{appendices}
\newpage
\section{Extended Data}

\begin{figure}[hbp]
\centering
\includegraphics[width=\textwidth]{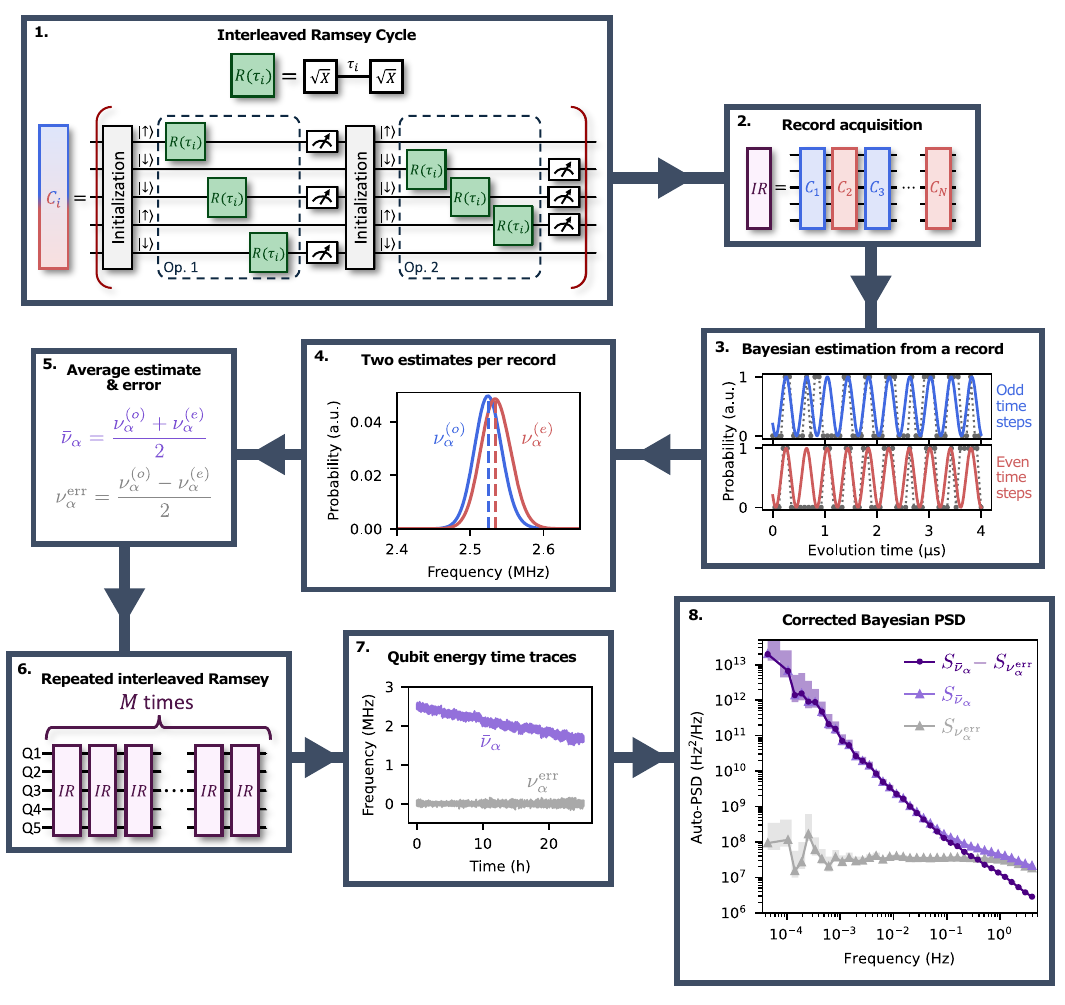}
\caption{\textbf{Interleaved Ramsey sequence and analysis workflow.} 
Schematic of the experimental protocol and data processing used to extract corrected PSDs for the qubits. 
Each of the five qubits is driven with an interleaved Ramsey sequence consisting of two operation steps, where the Ramsey block $R(\tau_i)$ with evolution time $\tau_i$ is applied to different subsets of qubits. 
The interleaved cycle is repeated with $\tau_i$ stepped from 0 to 4~$\mu$s in 0.04~$\mu$s increments, forming a record. 
For each qubit $\alpha$, we obtain two estimates of the qubit energy---$\nu_\alpha^{(o)}$ from odd $\tau_i$ and $\nu_\alpha^{(e)}$ from even $\tau_i$---from which we define an average estimate $\bar{\nu}_\alpha=(\nu_\alpha^{(o)}+\nu_\alpha^{(e)})/2$ and an error term $(\nu_\alpha^{(o)}-\nu_\alpha^{(e)})/2$. 
Records are acquired continuously over long durations (6--24 hours), yielding time traces of $\bar{\nu}_\alpha$ and the corresponding errors. 
Auto-PSDs are computed for both quantities, and the corrected auto-PSD of each qubit is defined as their difference. 
From the same $\bar{\nu}_\alpha$ traces, we also compute unnormalized cross-PSDs, which are unaffected by estimation errors. 
Normalized cross-PSDs are finally obtained by combining the unnormalized cross-PSDs with the corrected auto-PSDs.
}\label{exfig:sequence}
\end{figure}

\begin{figure}
\centering
\includegraphics[width=\textwidth]{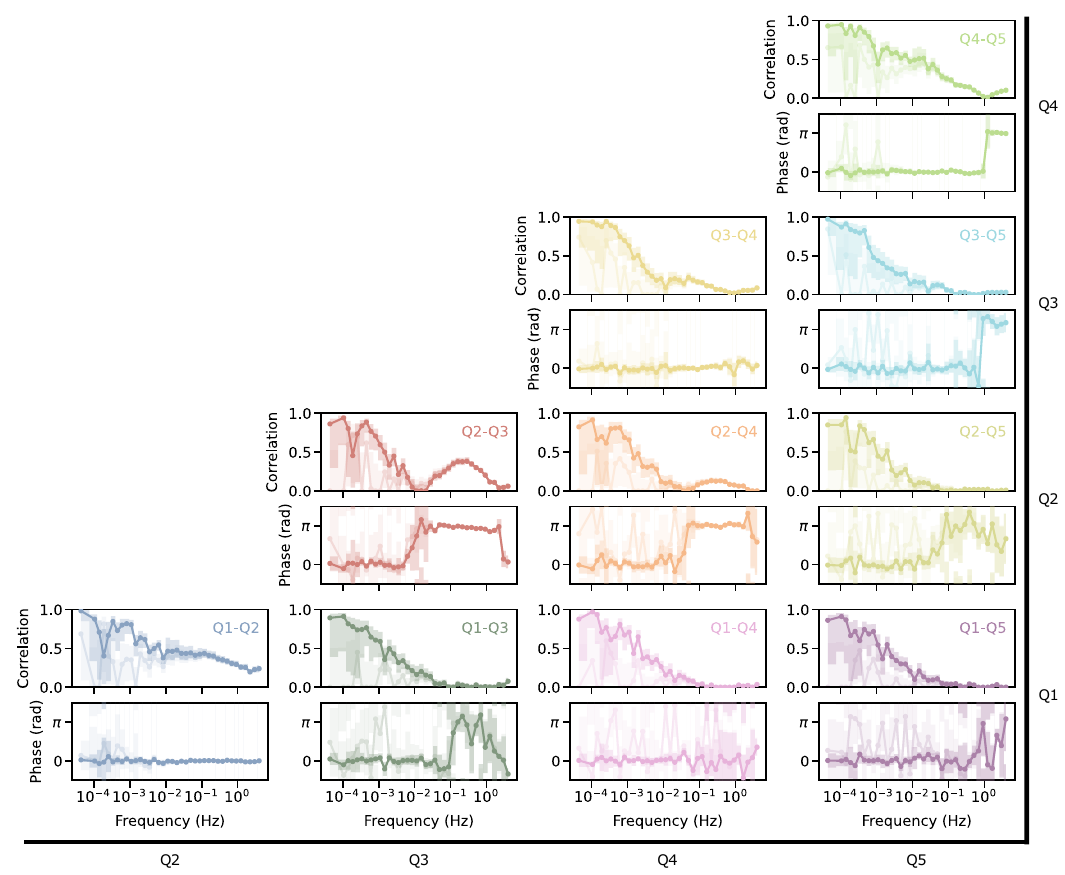}
\caption{\textbf{Correlation matrix.} 
Normalized cross-PSD amplitude and phase for all qubit pairs. 
Data points represent the maxima of the Bayesian posterior distributions for the PSD estimates, with error bars indicating 90\% confidence intervals. 
In each panel, lighter-colored points correspond to cross-PSDs computed from drift-corrected time traces.}\label{exfig:full_matrix}
\end{figure}

\begin{figure}
\centering
\includegraphics[scale=1]{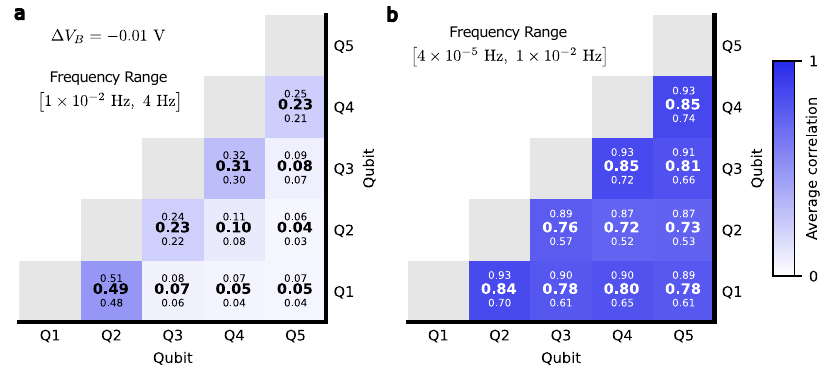}
\caption{\textbf{Correlation matrices in additional regimes.}
(a) Average correlation matrix for an intermediate barrier voltage $\Delta V_B=-0.01$~V between Q2 and Q3, shown for comparison with the two configurations in Fig.~\ref{fig:correlation_b}. The correlations interpolate smoothly between the two limits, demonstrating continuous electrostatic control. For this setting, the Q2--Q3 separation is $\sim123$~nm.
(b) Average correlation matrix evaluated in the frequency range $[4\times10^{-5}~\mathrm{Hz},\ 1\times10^{-2}~\text{Hz}]$, corresponding to the data in Extended Data Fig.~\ref{exfig:full_matrix} without drift removal. The global magnetic-field drift produces large, nearly uniform correlation amplitudes that are independent of interqubit distance. Such distance-independent correlations are particularly detrimental for quantum error correction and correspond to the fully correlated noise limit analyzed in Extended Data Fig.~\ref{exfig:limit_corr}.}\label{exfig:matrix_drift}
\end{figure}

\begin{figure}
\centering
\subfloat{\includegraphics[width=\textwidth]{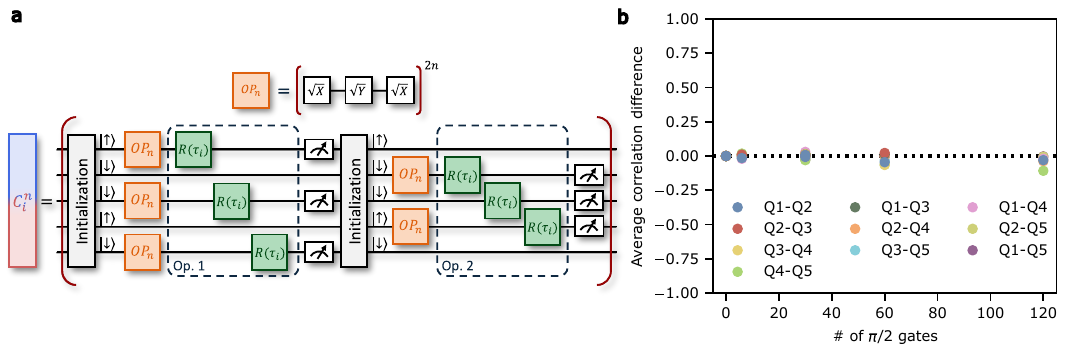}\label{exfig:operated_a}}
\subfloat{\label{exfig:operated_b}}
\caption{\textbf{Effect of gate operations on measured correlations.} 
To test whether noise correlations originate from our control pulses, we repeated the interleaved Ramsey sequence of Extended Data Fig.~\ref{exfig:sequence}, but preceded each Ramsey block with multiple gate operations that together implement an identity on the probed qubits. 
If the gates introduced additional correlated noise, the correlation amplitude should increase with the number of identity operations. 
Panel (a) illustrates the modified (``operated") Ramsey cycle, and panel (b) shows the difference in average correlations between the standard and operated sequences for all qubit pairs. 
No systematic increase in correlations is observed, which demonstrates that the measured correlations are intrinsic to the device environment rather than induced by our manipulations, and highlights the robustness of the correlation measurements. Notably, faster operation speeds require higher microwave amplitudes, which increase the delivered microwave power and can introduce additional noise in the system \cite{Wu2025}. In the present experiments, we operate in a regime where such effects are not observed.
}
\label{exfig:operated}
\end{figure}

\begin{figure}
\centering
\includegraphics[scale=1]{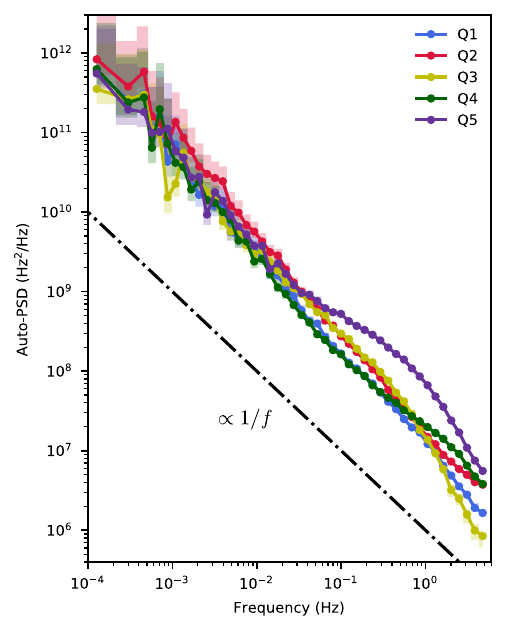}
\caption{\textbf{Auto-PSDs measured after a second cooldown.} 
Following a device warm-up, a second cooldown revealed different spectral characteristics. Shown here are the auto-PSDs of the qubits (colored regions indicate 90\% confidence intervals \cite{Gutierrez-Rubio2022}), which differ from those in Fig.~\ref{fig:auto-PSD}. In particular, Q5 now exhibits the strongest noise, with a Lorentzian feature similar to that previously observed in Q3 during the first cooldown. Although individual qubits display different spectral features compared to the first cooldown, the overall behavior and the presence of Lorentzian charge-noise signatures persist. The cross-PSDs also change, yet the overall statistical trend of correlation decay is preserved, as shown in Fig.~\ref{fig:scaling}.}\label{exfig:auto-PSD_2nd}
\end{figure}

\begin{figure}
\centering
\includegraphics[scale=1]{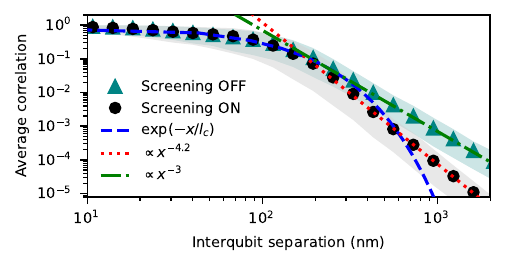}
\caption{\textbf{Long-range correlations decay of the TLF model.} 
Plot of the simulated average correlation as a function of interqubit distance, with (black circles) and without (teal triangles) screening. Shaded regions denote the 10--90 percentile range across 500 generated TLF ensembles. At short distances, both correlation decays are well described by an exponential (blue dashed line), crossing over to a polynomial tail at long distances. In the absence of screening, the average correlations decay as $\propto x^{-3}$ (green dashed-dotted line). Screening steepens this decay (dotted red line), yielding an approximate power-law dependence $\propto x^{-4.2}$.}
\label{exfig:model_log}
\end{figure}

\begin{figure}
\centering
\includegraphics[scale=1]{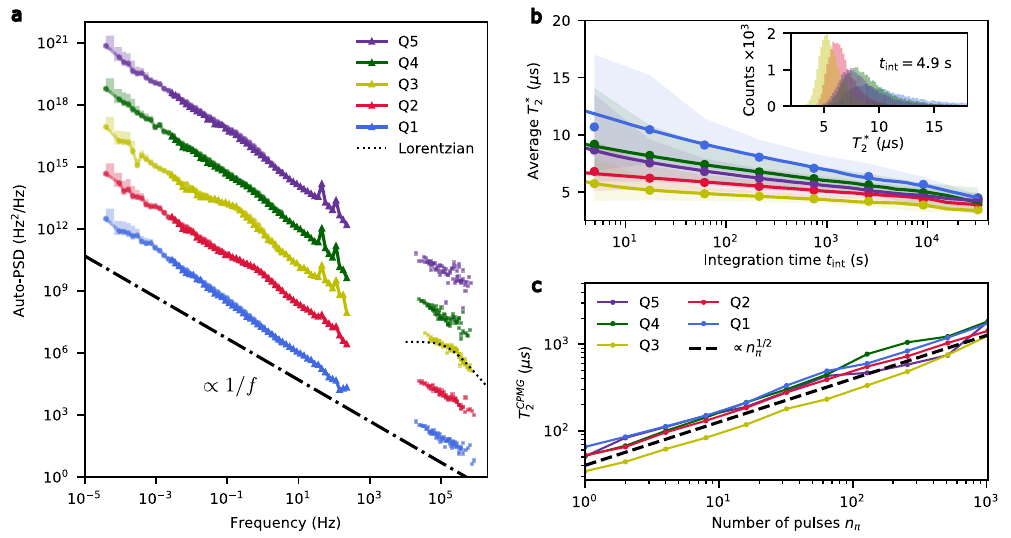}
\caption{\textbf{Wide range auto-PSDs and coherence times.} (a) The spectral range of the auto-power spectral densities (auto-PSDs) is extended by combining three complementary techniques: low-frequency spectroscopy from continuous Ramsey measurements (circles, as described in the main text), noise spectroscopy based on single-shot measurements from Ref.~\cite{Rojas-Arias2025b} (triangles), and CPMG dynamical decoupling \cite{Yoneda2018} (squares). The data were acquired under a voltage configuration different from that in Fig.~\ref{fig:auto-PSD}. All traces are vertically offset for clarity. The extended spectra reveal distinct peaks that appear across all qubits and increase in amplitude toward the right side of the array in Fig.~\ref{fig:device_d}. These peaks correspond to the fundamental and second harmonics of the 50~Hz electrical line frequency in Japan. The origin of the spatial bias in the peaks' amplitude remains unclear. Despite moderate scatter in the data, a Lorentzian-like feature (dotted line) emerges near $10^5$~Hz for Q3. The extended spectral range shows that the charge-noise mechanisms identified at low frequencies remain consistent across many decades of frequency, with the high-frequency spectrum reflecting the ensemble behavior of TLFs in the regime relevant for gate operations and QEC cycles. (b) Average coherence times as a function of the integration time, extracted from fits to decaying Ramsey oscillations. In this non-ergodic regime, the coherence time $T_2^*$ is a stochastic quantity whose average depends on the integration time \cite{Delbecq2016,Rojas-Arias2026a}. The inset shows histograms of $T_2^*$ for all qubits at a representative integration time. For each qubit, the mean values are plotted as circles in the main panel, with colored regions indicating the 5--95\% percentile ranges obtained from the histograms. The solid lines are calculated from the low-frequency auto-PSDs shown in (a). Specifically, the coherence time is evaluated as $T_2^*(t_{\mathrm{int}})=\bigl[2\pi^2 v(t_{\mathrm{int}})\bigr]^{-1/2}$, where the variance is $v(t)=2\int_{1/t}^{\infty}\mathrm{d}f\ S(f)$ and $S(f)$ is the two-sided auto-PSD of the corresponding qubit. The integral is split at the Nyquist frequency $f_{\mathrm{Nyq}}=1/(2t_s)$, where $1/t_s$ is the sampling rate of the qubit-energy time traces. The low-frequency contribution, $v(t)=2\int_{1/t}^{f_{\mathrm{Nyq}}}\mathrm{d}f\ S(f)$ is obtained directly from the measured auto-PSDs extracted from continuous Ramsey measurements (circles in (a)). The high-frequency contribution is treated as a fitting constant [in fitting the data (points) in panel (b)], due to the limited spectral bandwidth accessible experimentally. The resulting solid lines agree well with the average $T_2^*$ extracted from Ramsey decay fits, demonstrating the quantitative connection between the noise spectrum and the coherence time. Similarly, the integrated auto-PSD can be related to gate fidelities as done in Ref.~\cite{Wu2025}. (c) Coherence times as a function of the number of $\pi$ pulses $n_\pi$ in the dynamical-decoupling sequence used to extract the high-frequency spectra (squares in (a)). All coherence times scale as $T_2\propto n_\pi^{1/2}$, consistent with a $1/f$ noise spectrum.}
\label{exfig:wide}
\end{figure}

\begin{figure}
\centering
\includegraphics[scale=1]{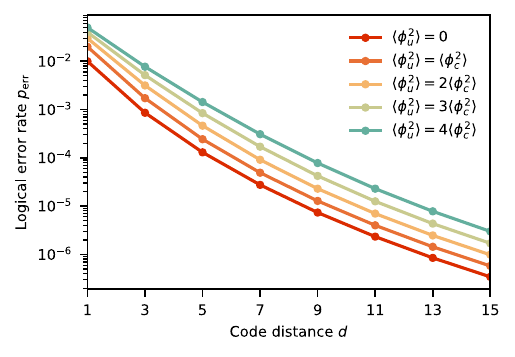}
\caption{\textbf{Lower bound of logical error rate imposed by perfectly correlated noise.} A global noise source that induces the same random phase on all qubits with variance $\langle\phi_c^2\rangle$ sets a fundamental lower bound on the logical error rate of the repetition code. In this case, all cross-PSDs are $c_{i,j}=1$ and the single-qubit error probability is $p\approx \langle\phi_c^2\rangle/4$ (Eq.~\eqref{eq:single_error}). We take a representative single-qubit error rate of $p=10^{-2}$. When noise is fully correlated, phase fluctuations act collectively on all qubits, increasing the likelihood of simultaneous errors that the repetition code cannot correct. Adding uncorrelated noise with variance $\langle\phi_u^2\rangle$ increases the single-qubit error probability to $p'\approx(\langle\phi_c^2\rangle+\langle\phi_u^2\rangle)/4$, while reducing the correlation amplitude to $c'_{i,j}=\langle\phi_c^2\rangle/(\langle\phi_c^2\rangle+\langle\phi_u^2\rangle)$ (Eq.~\eqref{eq:cross_identical}). The behavior shown in this figure demonstrates that this reduction in correlation is counteracted by the increased single-qubit error probability, such that all curves with added uncorrelated noise lie above the line defined by the perfectly correlated case. Consequently, perfectly correlated noise, such as global magnetic field drift, sets a hard lower bound on the achievable logical error rate in the repetition code.}\label{exfig:limit_corr}
\end{figure}

\begin{figure}
\centering
\includegraphics[scale=1]{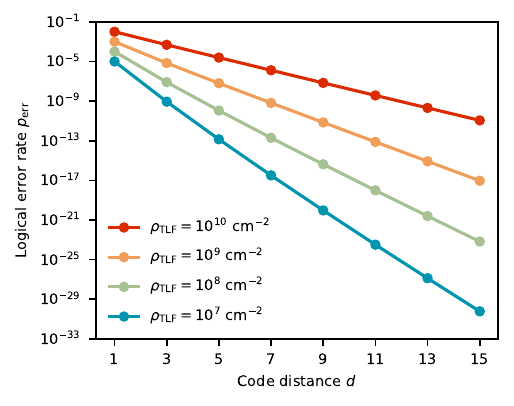}
\caption{\textbf{Repetition code performance for different TLF densities.}
The spatial decay length of noise correlations is primarily set by the density of TLFs \cite{Rojas-Arias2023,Boter2020}. At low TLF densities, multiple qubits couple to the same small set of fluctuators, resulting in stronger and longer-range correlations. At high densities, each qubit is dominated by nearby fluctuators, which overshaddow more distant ones, leading to weaker correlations and an approximately local noise environment. This behavior is illustrated by the dotted curve in Fig.~\ref{fig:scaling}. On the other hand, reducing the TLF density simultaneously lowers the overall noise amplitude and therefore the physical phase-flip error probability. Similarly as in Fig.~\ref{exfig:limit_corr}, it raises the question of whether increased correlations at low densities could outweigh the benefit of reduced physical error rates and lead to higher logical error rates. To address this, we compute the logical error rate of the repetition code for different TLF densities. For each density, we generate 500 independent TLF ensembles to extract the corresponding correlation-versus-distance profiles, which are then used as input for the repetition-code calculation. As a reference point, we assign a physical error rate $p=10^{-2}$ at a density $\rho_{\mathrm{TLF}}=10^{10}\ \mathrm{cm}^{-2}$ and scale the physical error rate as $p\propto\rho_{\mathrm{TLF}}$. This scaling follows from $p\simeq\langle\phi^2\rangle/4\propto S(f)$ (Eqs.~\eqref{eq:auto_identical} and \eqref{eq:single_error}) and from the relation $S(f)\propto\rho_{\mathrm{TLF}}$. Despite the increase in correlation length at lower densities, we find that the reduction in physical error rate dominates, leading to a monotonic decrease of the logical error rate with decreasing TLF density. Thus, the conclusion from Figs.~\ref{exfig:limit_corr} and \ref{exfig:repetition_vs_density} is that removing noise is always beneficial, irrespective of what it implies for the degree of noise correlations, a natural conclusion.
}\label{exfig:repetition_vs_density}
\end{figure}

\begin{figure}
\centering
\includegraphics[scale=1]{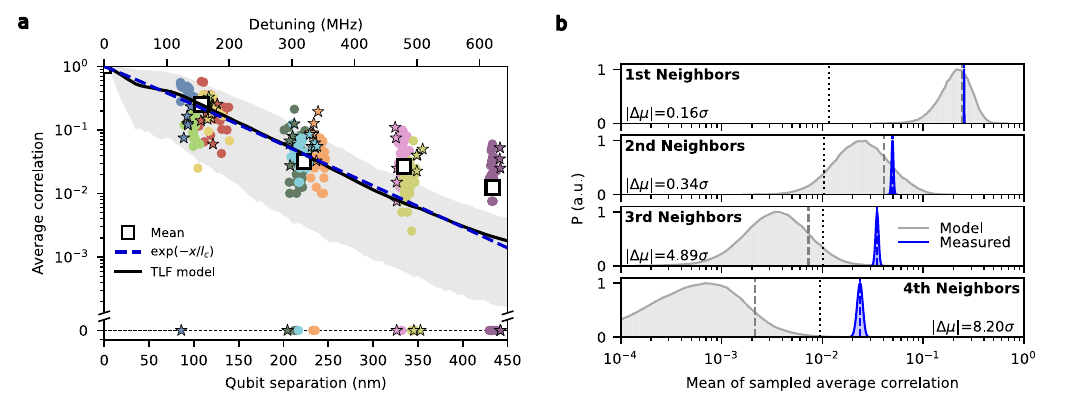}
\caption{\textbf{Remnant correlation at large interqubit separation.}
(a) Same data as Fig.~\ref{fig:scaling} (same symbol conventions), shown on a logarithmic vertical scale. Each point corresponds to the peak of the Bayesian posterior distribution of the correlation obtained for a single measurement, as described in Methods. At large interqubit separations the posteriors are frequently peaked at zero, reflecting the finite resolution of the estimation procedure due to finite data size. While the TLF model reproduces both the decay and variability of the average correlation for first- and second-neighbor separations, the measured correlations for third and fourth neighbors exhibit an apparent saturation.
(b) To assess whether this deviation is statistically significant and not an artifact of limited statistics, we propagate the full Bayesian posteriors to the level of the mean correlation at each neighboring level. For a given neighboring level, we draw one sample from each posterior distribution and compute the mean of the sampled values. Repeating this procedure $N_\mathrm{MC}=2\times10^5$ times yields a distribution of mean correlations (blue histogram). A solid blue curve shows the kernel density estimate (KDE) of this distribution. The dashed blue vertical line indicates its mean, denoted $\mu_\mathrm{meas}$.
The finite resolution limit of the Bayesian estimator for each individual measurement is approximately $1/\sqrt{M_\mathrm{eff}}$ \cite{Gutierrez-Rubio2022}, where $M_\mathrm{eff}$ is the effective number of independent batches (typically of order $10^4$; see Methods). The average resolution limit for each neighboring level is shown as a vertical black dotted line. In all cases the sampled mean correlation exceeds this scale, confirming that the observed means are not set by the single-measurement resolution floor.
For comparison, we construct the corresponding distribution of mean correlations predicted by the TLF model at the representative interqubit separation for each neighbor level. Each synthetic mean is obtained by averaging $N_\mathrm{model}$ independent draws from the model distribution. In a conservative scenario, we take $N_\mathrm{model}=2N_\mathrm{neigh}$, assuming that only measurements from different cooldowns and different qubit pairs probe independent TLF ensembles. The resulting model distribution is shown as a gray histogram with a solid gray KDE curve. The dashed gray vertical line indicates its mean, denoted $\mu_\mathrm{model}$.
For first and second neighbors, the measured and model mean distributions overlap substantially, indicating quantitative agreement. In contrast, for third and fourth neighbors the overlap is negligible. We quantify the separation between measurement and model by $\Delta\mu = \mu_\mathrm{meas}-\mu_\mathrm{model}$ expressed in units of 
$\sigma=\sqrt{\sigma_\mathrm{meas}^2+\sigma_\mathrm{model}^2}$, where $\sigma_\mathrm{meas}$ and $\sigma_\mathrm{model}$ are the standard deviations of the measured and model distributions, respectively. The deviations correspond to $4.89\sigma$ for third neighbors and $8.20\sigma$ for fourth neighbors, demonstrating statistically significant remnant correlations beyond the TLF-only model. Possible origins include incomplete drift removal or additional spatially correlated noise sources, such as voltage fluctuations in shared gate electrodes. Nevertheless, this residual correlation has a negligible impact on the QEC results in Fig.~\ref{fig:qec}, altering the predicted logical error rates by less than 7\%.
}\label{exfig:deviation}
\end{figure}

\end{appendices}

\clearpage

\bibliography{references.bib}

\end{document}